\documentclass[preprint,12pt]{elsarticle}

\usepackage{amssymb,amsmath,amsthm}
\usepackage{ dsfont }
\usepackage{color}

\newcommand{\Prob} {\mbox{$\rm{Prob}$\,}}

\journal{Theoretical Population Biology}

\begin{document}

\begin{frontmatter}



\title{An approximate stationary solution for multi-allele neutral diffusion \textcolor{black}{with low mutation rates}.}

\author[label1,label2]{Conrad J.\ Burden}
\address[label1]{Mathematical Sciences Institute, Australian National University, Canberra, Australia}
\address[label2]{Research School of Biology, Australian National University, Canberra, Australia}
\ead{conrad.burden@anu.edu.au}
\author[label1]{Yurong Tang}
\ead{yurong.tang@anu.edu.au}

\begin{abstract}
We address the problem of determining the stationary distribution of the multi-allelic, neutral-evolution Wright-Fisher model in the diffusion limit.  A full solution to this problem for an arbitrary $K \times K$ mutation rate matrix involves solving for the stationary solution of a forward Kolmogorov equation over a $(K - 1)$-dimensional simplex, and remains intractable.  In most practical situations mutations rates are slow on the scale of the diffusion limit and the solution is heavily concentrated on the corners and edges of the simplex.  In this paper we present a practical approximate solution for slow mutation rates in the form of a set of line densities along the edges of the simplex.  The method of solution relies on parameterising the general non-reversible rate matrix as the sum of a reversible part and a set of $(K - 1)(K - 2)/2$ independent terms corresponding to fluxes of probability along closed paths around faces of the simplex.  The solution is potentially a first step in estimating 
\textcolor{black}{
non-reversible
}
evolutionary rate matrices from observed allele frequency spectra.  
\end{abstract}


\end{frontmatter}



\section{Introduction}
\label{sec:Introduction}

The rapidly reducing cost of high throughput sequencing now allows for the acquisition of genome-wide data 
\textcolor{black}{
for detecting nucleotide allele frequencies extracted from multiple alignments within a population across large numbers of 
genomic sites~\cite{pool2010population}.  
}
The existence of such data raises the possibility of estimating not only specific mutation rates, 
but complete evolutionary rate matrices from the current observed state of allele frequencies with the genome.  

In a recent paper Vogl~\cite{vogl2014estimating} has developed a general algorithm and, in the limit of slow scaled mutation rates, a maximum likelihood 
estimate, of the two parameters defining the scaled instantaneous rate matrix for the case of bi-allelic neutral evolution.  The estimator is similar in style to  
Watterson's estimator for the infinite allele case~\cite{watterson1975number}, and assumes the data to consist of a site-frequency spectrum 
(or allele-frequency spectrum) obtained from genotyping a finite number of individuals at a relatively large number of independent sites whose evolution 
is subject only to genetic drift and identical-rate mutations.  It is derived by assuming the data has a beta-binomial distribution as a result of being sampled 
from the well-known beta-distribution solution to the diffusion limit of the neutral Wright-Fisher model~\cite{wright1931evolution}.   
\textcolor{black}{
The method is extended to include selection and the analysis of the low mutation rate limit developed further by Vogl and Berman in 
\cite{vogl2015inference}.  
}

A necessary first step in generalising the Vogl estimator to the multi-allele case, and in particular to the 4-allele case relevant to genomic rate matrices, 
is the generalisation of Wright's stationary beta distribution to higher dimensions.   This involves finding a stationary solution to the multi-allelic forward  
Kolmogorov equation (see Eq.~(\ref{generalPDE}) in the next section).   There is no known general solution to this partial differential equation for 
an arbitrary instantaneous rate matrix.  

However, physical mutation rates are extremely slow on the scale relevant to the diffusion limit, and therefore we argue that for practical purposes 
it is not necessary to solve the forward  Kolmogorov equation in its entirety over the full volume of the 3-dimensional simplex on which its solution 
is defined.  Consider for instance the numerical stationary solution to the discrete Wright-Fisher defined by Eqs.~(\ref{eq:FullPij}) to (\ref{QToU}) below, 
shown in Fig.~\ref{fig:Tetrahedron}.  For the purposes of illustration we have simulated this solution 
using the popular Hasegawa-Kishino-Yano matrix (HKY85)~\cite{hasegawa1985dating} with a small population in order to render the simulation 
numerically tractable, and mutation rates which are unrealistically high by at least two orders of magnitude to enable the distribution to be visible over 
the entire simplex on the scale of the plot.  

\begin{figure}[!t]
  \centering
   \includegraphics[width=0.6\textwidth]{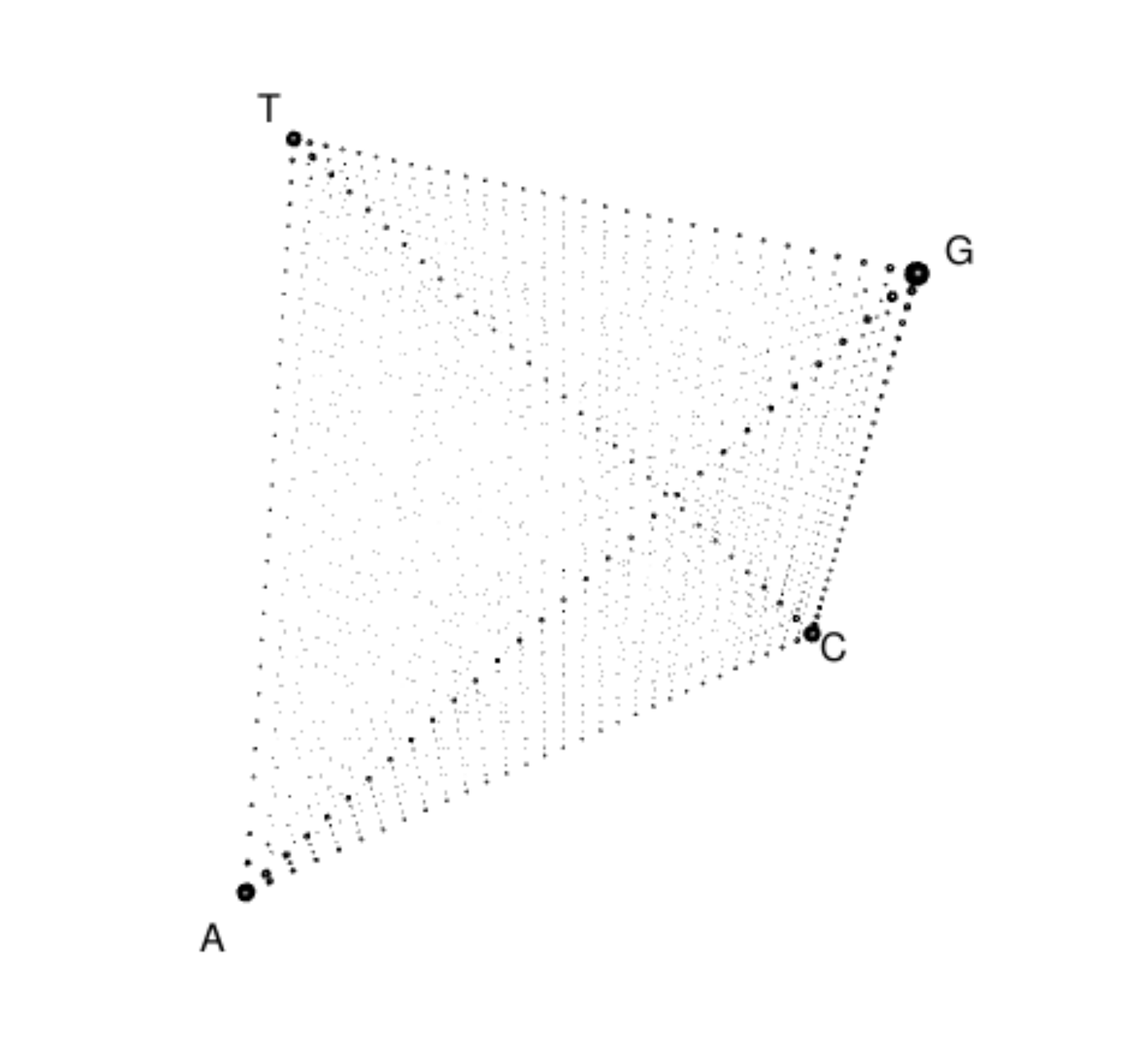} 
  \caption{Stationary distribution of allele frequencies for the HKY85 model for a haploid population of size $N = 30$ with parameters 
  $\alpha = 0.2$, $\beta = 0.1$, $\pi_A = \pi_T = 0.2$ and $\pi_C = \pi_G = 0.3$, using the parameterisation defined in ref.~\cite{hasegawa1985dating}.  
  The corners labelled $A$, $C$, $G$ and $T$ correspond to allele frequencies $\mathbf{i} = (N, 0, 0, 0)$, $(0, N, 0, 0)$, 
  $(0, 0, N, 0)$ and $(0, 0, 0, N)$ respectively, and the volume of the sphere at each coordinate point is proportional to the probability mass function.}
  \label{fig:Tetrahedron}
\end{figure}

The distribution is clearly dominated by the corners of the tetrahedron, indicating that the majority of genomic sites are not polymorphisms (SNPs).  
\textcolor{black}{
This effect is explained in \cite{vogl2015inference} in the context of the 2-allele Moran model as a strong dominance of genetic drift over mutaions 
for polymorphic sites.  
}
Most of the remaining support of the distribution lies on the edges of the tetrahedron, which correspond to 2-allele SNPs.
The interiors of the four faces, corresponding to 3-allele SNPs, and the interior volume of the tetrahedron, corresponding to 4-allele SNPs, account for 
only a small fraction of the total probability.  Consistent with observation of the human genome~\cite{hodgkinson2010human,cao2015analysis,phillips2015tetra}, 
the multi-allele neutral Wright-Fisher model predicts that 3- and 4-allele SNPs are extremely rare when scaled mutation rates are low.  In fact, when 
tri-allelic SNPs are observed, the least frequent allele is generally observed in only 1 or 2 percent of the population 
(see Table~S1 of \cite{hodgkinson2010human}), corresponding to points very close to 
an edge of the tetrahedron.  

Below we present an approximate solution to the multi-allelic forward Kolmogorov equation in the form of a set of line densities defined on the edges 
of the solution simplex for the general case of $K$ alleles.  The basis of our solution is a novel parameterisation of the most general form of the instantaneous 
rate matrix $Q$, subject only to the constraints that its off-diagonal elements be non-negative and that its rows sum to zero.  The parameterisation consists 
of writing $Q$ as the sum of a time-reversible part~\cite{tavare1986some} plus a non-reversible part parametrised by $(K - 1)(K - 2)/2$ `probability fluxes' 
corresponding to a set of independent closed triangular paths following edges of the solution simplex.  The assumption that rate matrices are 
reversible is popular in the phylogenetics literature because the pulley principle~\cite{felsenstein1981evolutionary} simplifies calculations.  However there is 
no biochemical justification for this assumption.  We find that in the limit of low mutation rates, and if neutral evolution is assumed, asymmetry in the allele 
frequency spectrum along edges of the solution simplex can only be explained by the non-reversible part of $Q$.  Equivalently, if $Q$ is reversible, 
the allele frequency spectrum is symmetric along each edge. 

The structure of this paper is as follows.  
Section~\ref{sec:Review} contains a review the multi-allelic neutral Wright-Fisher model and sets out the statement of the problem.  
Section~\ref{sec:K2Case} reviews the $K = 2$ solution to the forward Kolmogorov equation with a focus on non-standard boundary conditions.  
Sections~\ref{sec:K3Case}, \ref{sec:K4Case} and \ref{sec:ArbitraryK} contain our approximate solutions for the $K = 3$, $4$ and arbitrary $K$ cases respectively.  
\textcolor{black}{
Section~\ref{sec:StrandSymmetry} discusses the strand-symmetric case.   
Conclusions are summarised in Section~\ref{sec:Conclusions}.   
\ref{sec:StationaryBoundary} is devoted to deriving the asymptotic behaviour of the solution 
to Eq.~(\ref{generalPDE}) near the simplex boundary in the limit of low mutation rates.  
}
\ref{sec:Equiv2} is devoted to technical details of obtaining marginal distributions 
of the stationary $K$-allele solution in terms of effective 2-allele models.  


\section{Review of the multi-allelic neutral Wright-Fisher model}
\label{sec:Review}

We consider the neutral evolution Wright-Fisher model for $K$ alleles, labelled $A_1$ \ldots $A_K$ (see, for example, Section~4.1 of ref.~\cite{etheridge2011some}).  
Given a haploid population of size $N$ (or diploid population of size $N/2$), 
let the number of individuals of type $A_a$ at time step $\tau$ be $Y_a(\tau)$ for discrete times $\tau = 0, 1, 2, \ldots$.  
Also, let $u_{ab}$ be the probability of an individual making a transition from $A_a$ to $A_b$ in a single time step, where $u_{ab} \ge 0$ and 
$\sum_{b = 1}^K u_{ab} = 1$.  Writing $\mathbf Y(\tau) = (Y_1(\tau), \ldots Y_K(\tau)))$, the multi-allele neutral Wright-Fisher model is defined by the transition 
matrix from an allele frequency $\mathbf i = (i_1 , \ldots, i_K)$ to an allele frequency $\mathbf j = (j_1 , \ldots, j_K)$ in the population given by  
\begin{equation}
\Prob(\mathbf Y (\tau + 1) = \mathbf j | \mathbf Y (\tau) = \mathbf i) = 
            \frac{N!}{\prod_{a = 1}^K j_a!} \prod_{a=1}^K{\psi(\mathbf{i}, a})^{j_a}  ,  \label{eq:FullPij}
\end{equation}
where $\sum_{a = 1}^K i_a = \sum_{a = 1}^K j_a = N$, and 
\begin{equation}
\psi(\mathbf{i}, a) = \frac{i_a}{N}  \left(1 - \sum_{b \ne a} u_{ab} \right) + \sum_{b \ne a} \frac{i_b}{N} u_{ba} = \sum_{b = 1}^K \frac{i_b}{N} u_{ba}.  \label{psiDef}
\end{equation}
\textcolor{black}{
This transition matrix defines a finite state Markov chain with a state space of dimension ${N + K - 1}\choose{K - 1}$.  The distribution in Eq.~(\ref{eq:FullPij}) 
is a multinomial distribution with probabilities $\psi(\mathbf{i}, a)$.  
}

The usual diffusion limit is obtained by defining random variables $X_a(t) = Y_a(\tau)/N$ equal to the relative proportion of type-$A_a$ 
alleles within the population at continuous time $t = \tau/N$.  The limit $N \rightarrow \infty$ and $u_{ab} \rightarrow 0$ for $a \ne b$ is taken in such a 
way that the $K \times K$ instantaneous rate matrix $Q$, whose elements are defined by 
\begin{equation}
Q_{ab} = N (u_{ab} - \delta_{ab}),	\label{QToU}
\end{equation}
remains finite.  Here $\delta_{ab}$ is the Kronecker delta, equal to 1 if $a = b$ and 0 otherwise.  This limit gives the forward Kolmogorov equation 
\begin{equation}
\frac{\partial f}{\partial t} = - \sum_{a = 1}^{K - 1} \frac{\partial}{\partial x_a}  \sum_{b = 1}^K x_b Q_{b a} f  
	+ \frac{1}{2} \sum_{a, b = 1}^{K - 1} \frac{\partial^2}{\partial x_a \partial x_b} \left\{  (\delta_{a b} x_a - x_a x_b) f \right\}, 
										\label{generalPDE}
\end{equation}
for the density function $f(x_1, \ldots x_{K-1}; t)$ of the vector of continuous random variables $X_1(t), \ldots X_{K - 1}(t)$.  The function $f$ is defined over the simplex \begin{equation}
\left\{ (x_1, \ldots, x_{K - 1}) : x_1, \ldots, x_{K - 1} \ge 0, \, \sum_{a = 1}^{K - 1} x_a \le 1\right\}.  \label{simplex}
\end{equation} 
For notational convenience, we have defined $x_K = 1 - \sum_{a = 1}^{K - 1} x_a$ in Eq.~(\ref{generalPDE}), and also in Eq.~(\ref{DirSolution}) below.  
\textcolor{black}{
For details of the derivation of Eq.~(\ref{generalPDE}) see Lemma~4.2 of \cite{etheridge2011some} or Eq.~(5.125) of \cite{Ewens:2004kx}. 
}

The stationary distribution $f(x_1, \ldots x_{K-1})$ to the forward Kolmogorov equation is obtained by setting $\partial f/\partial t$ to zero.  
The general solution to this problem for an arbitrary rate matrix $Q$ is unknown.  However, it is known for the special case in which the elements of the rate 
matrix take the form $Q_{ab} = Q_b$ (independent of $a$), in which case the solution is a Dirichlet distribution~\cite{wright1969evolution,tier1978tri,griffiths1979transition}, 
namely
\begin{equation}
f(x_1, \ldots x_{K-1}) = \Gamma\left(2\sum_{a = 1}^K Q_a\right) \prod_{a = 1}^K \frac{x_a^{2Q_a - 1}}{\Gamma(2Q_a)}.  \label{DirSolution}
\end{equation}  
Rate matrices of this form constitute a subset of the set of general time-reversible rate matrices.  

Our aim is to explore the stationary solution in the limit of slow but otherwise arbitrary mutation rates, that is for 
\textcolor{black}{
rate matrices $Q$ constrained only by the requirements that 
}
$0 \le Q_{a b} <<1$ for $a \ne b$ and 
$\sum_{b = 1}^K Q_{ab} = 0$.  Our analysis is based on two 
\textcolor{black}{
observations.  First, in the limit $Q_{ab} \rightarrow 0$, the probability distribution 
$f$ is concentrated along the edges of the simplex, and therefore the solution to Eq.~(\ref{generalPDE}) can be represented accurately 
as a set of line densities defined along the edges of the simplex, as demonstrated in \ref{sec:StationaryBoundary}.   
}
Second, we assume that marginal distributions of the stationary distribution to Eq.~(\ref{generalPDE}) 
corresponding to partitioning the set of alleles into two distinct subsets (as described in \ref{sec:Equiv2}) are Wright's \cite{wright1931evolution} 
well-known beta function solutions to the $K = 2$ case.  


\section{$K = 2$ case}
\label{sec:K2Case}

The solution to the $K = 2$ case is of course well known 
\textcolor{black}{
(see for example Section~5.6 of \cite{Ewens:2004kx}).  
}
However we summarise the solution in order to establish notation and to draw attention to 
properties associated with non-standard boundary conditions.  
Set $x$ equal to the proportion of $A_1$ alleles and $(1 - x)$ equal to the proportion of $A_2$ alleles present in a population.  
Setting $K = 2$ in Eq.(\ref{generalPDE}) and integrating once yields 
\begin{equation}
\{Q_{12}x - Q_{21}(1 - x)\} f(x) + \frac{1}{2} \frac{d}{dx} \{x(1 - x) f(x) \} = \Phi, \label{deForKEquals2}
\end{equation}
where the constant of integration $\Phi$ represents a flux of probability per unit time across any point in the interval $[0, 1]$.  
\textcolor{black}{
This flux is generally set equal to zero since, when only 2 alleles are present,  probability cannot flow across the boundary points at $x = 0$ and $1$.  
However it will be necessary in subsequent sections to consider the solution to Eq.~(\ref{deForKEquals2}) when $\Phi$ is non-zero.  
}

The most general solution can be written in a form which is symmetric with respect to the two alleles $A_1$ and $A_2$ as 
\begin{eqnarray}
\lefteqn{f(x; C, \Phi) = } \nonumber \\
& & \left[C + \{B(x; 1 - 2Q_{21}, 1 - 2Q_{12}) -  B(1 - x; 1 - 2Q_{12}, 1 - 2Q_{21}) \} \Phi \right] x^{2Q_{21} -1} (1 - x)^{2Q_{12} -1}, \nonumber\\
							\label{dEquals2Soln}
\end{eqnarray}
where $C$ is a constant of integration and
\begin{equation}
B(x; \alpha, \beta) = \int_0^x \xi^{\alpha -1} (1 - \xi)^{\beta - 1} d\xi,
\end{equation}
is the incomplete beta function, defined for any two parameters $\alpha$ and $\beta$.  The usual solution, first quoted by Wright~\cite{wright1931evolution}, 
is obtained by setting $\Phi = 0$ and $C = 1/B(2Q_{21}, 2Q_{12})$, where $B(\alpha, \beta) = B(1; \alpha, \beta)$ is the complete beta function.  

Equation~(\ref{dEquals2Soln}) takes a particularly simple form in the limit $0 \le Q_{12}, Q_{21} < \epsilon << 1$ provided $x$ is not close to the boundaries 0 or 1.  
In this case we have 
\begin{eqnarray}
x^{2Q_{21} -1} (1 - x)^{2Q_{12} -1} & = & \frac{e^{2Q_{21}\log x + 2Q_{12}\log(1 - x)}}{x(1 - x)} \nonumber \\
		& = &\frac{1}{x(1 - x)}\{1 + O(\epsilon \left|\log x + \log(1 - x)\right|)\},
\end{eqnarray} 
as $\epsilon \rightarrow 0$.  Furthermore
\begin{eqnarray}
B(x; 1 - 2Q_{21}, 1 - 2Q_{12}) & = & \int_0^x \xi^{-2Q_{21}} (1 - \xi)^{-2Q_{12}} d\xi \nonumber \\
& = & \int_0^x \xi^{-2Q_{21}}  d\xi \{1 + O(\epsilon)\} \nonumber \\
& = & x^{1 - 2Q_{21}} \{1 + O(\epsilon)\} \nonumber \\
& = & x\{1 + O(\epsilon (1 + \left|\log x\right|)\} , 
\end{eqnarray}
and similarly
\begin{equation}
B(1 - x; 1 - 2Q_{12}, 1 - 2Q_{21})  = (1 - x)\{1 + O(\epsilon (1 + \left|\log (1 - x)\right|)\},.
\end{equation}
Thus 
\begin{equation}
f(x; C, \Phi) \approx \frac{C}{x(1 - x)} - \Phi\left(\frac{1}{x} - \frac{1}{1 - x} \right),  \label{approx2AlleleSoln}
\end{equation}
provided $\max(Q_{12}, Q_{21}) \times \left|\log x + \log(1 - x)\right| << 1$.  
\textcolor{black}{
Note that the constant $C$ depends on the $2 \times 2$ rate matrix $Q$, but the flux $\Phi$ arises as a constant of integration and is 
independent of $Q$, which is necessarily reversible when $K = 2$.    Note also that the first term in Eq.~(\ref{approx2AlleleSoln}) 
is encapsulated within the boundary-selection-mutation model of Vogl and Berman~\cite{vogl2015inference}.  In this model 
the distribution is accounted for by only genetic drift and selection in the region $x \in [1/N, 1 - 1/N]$ and the local effects of mutation are 
safely ignored.  
}


\section{$K = 3$ case}
\label{sec:K3Case}

Consider the most general form of a $3 \times 3$ instantaneous rate matrix $Q$, the only restrictions on its elements being that $Q_{a b} \ge 0$ for $a \ne b$ and 
$\sum_{b = 1}^3 Q_{a b} = 0$.  It will prove convenient in what follows to parameterise $Q$ as 
\begin{equation}
\begin{split}
Q & =  \left(
\begin{array}{ccc}
-(\alpha_{12} \pi_2 + \alpha_{13} \pi_3)	&	\alpha_{12} \pi_2	&	\alpha_{13} \pi_3	\\
\alpha_{21} \pi_1	&	-(\alpha_{21} \pi_1 + \alpha_{23} \pi_3)	&	\alpha_{23} \pi_3	\\
\alpha_{31} \pi_1	&	\alpha_{32} \pi_2	&	-(\alpha_{31} \pi_1 + \alpha_{32} \pi_2)	
\end{array} 
			\right)     \\
	&  \qquad \qquad + 
\frac{\Phi}{2} \left( 
\begin{array}{ccc}
0	&	1/\pi_1	&	-1/\pi_1	\\
-1/\pi_2	&	0	&	1/\pi_2	\\
1/\pi_3	&	-1/\pi_3	&	0	
\end{array} 
			\right)   \\  \\
& =  Q^{\rm GTR} + Q^{\rm flux}, 	
\end{split}						\label{general3By3Q}
\end{equation}
subject to the constraints $\alpha_{ab} = \alpha_{ba} \ge 0$, $\pi_a \ge 0$ and $\sum_{a = 1}^3 \pi_a = 1$.  The requirement that the off-diagonal elements of $Q$ 
are non-negative implies the further constraint that 
\begin{equation}
 \left| \Phi \right| \le \min_{1 \le a < b \le 3} (2\alpha_{ab} \pi_ a\pi_b).	\label{phiRestriction}
\end{equation}

The first term, $Q^{\rm GTR}$, is the general time-reversible rate matrix~\cite{lanave1984new,tavare1986some}, with stationary distribution 
$\pi^T = (\pi_1, \pi_2, \pi_3)$ satisfying 
$\pi^T Q = \pi^T Q^{\rm GTR}  = 0$.  The defining property of a time-reversible rate matrix, namely that its elements $Q^{\rm GTR}_{ab}$ satisfy 
\begin{equation}
\pi_a Q^{\rm GTR}_{ab} = \pi_b Q^{\rm GTR}_{ba}, \quad a, b = 1, 2, 3
\end{equation}
implies that, for a Markov chain in its stationary state, transitions from allele $A_a$ to allele $A_b$ occur with the same frequency as transitions from 
allele $A_b$ to allele $A_a$.  The second term, $Q^{\rm flux}$, represents a net rate $\Phi$ of transitions around the closed loop $A_1 \rightarrow A_2 \rightarrow A_3 \rightarrow A_1$.   
\textcolor{black}{
The factor of $1/2$ ensures that, when the stationary state is achieved, the number of transitions from $A_a$ to $A_b$ minus the number 
of transitions from $A_b$ to $A_a$ per unit time, namely $\pi_a Q_{ab} - \pi_b Q_{ba}$, is $\Phi$.  
}
As required for a $3 \times 3$ rate matrix the total number of independent parameters is six: $\alpha_{12}$, $\alpha_{13}$ and $\alpha_{23}$, 
any two of $\pi_1$, $\pi_2$ and $\pi_3$, and $\Phi$.  

\begin{figure}[t]
   \centering
   \includegraphics[width=0.9\textwidth]{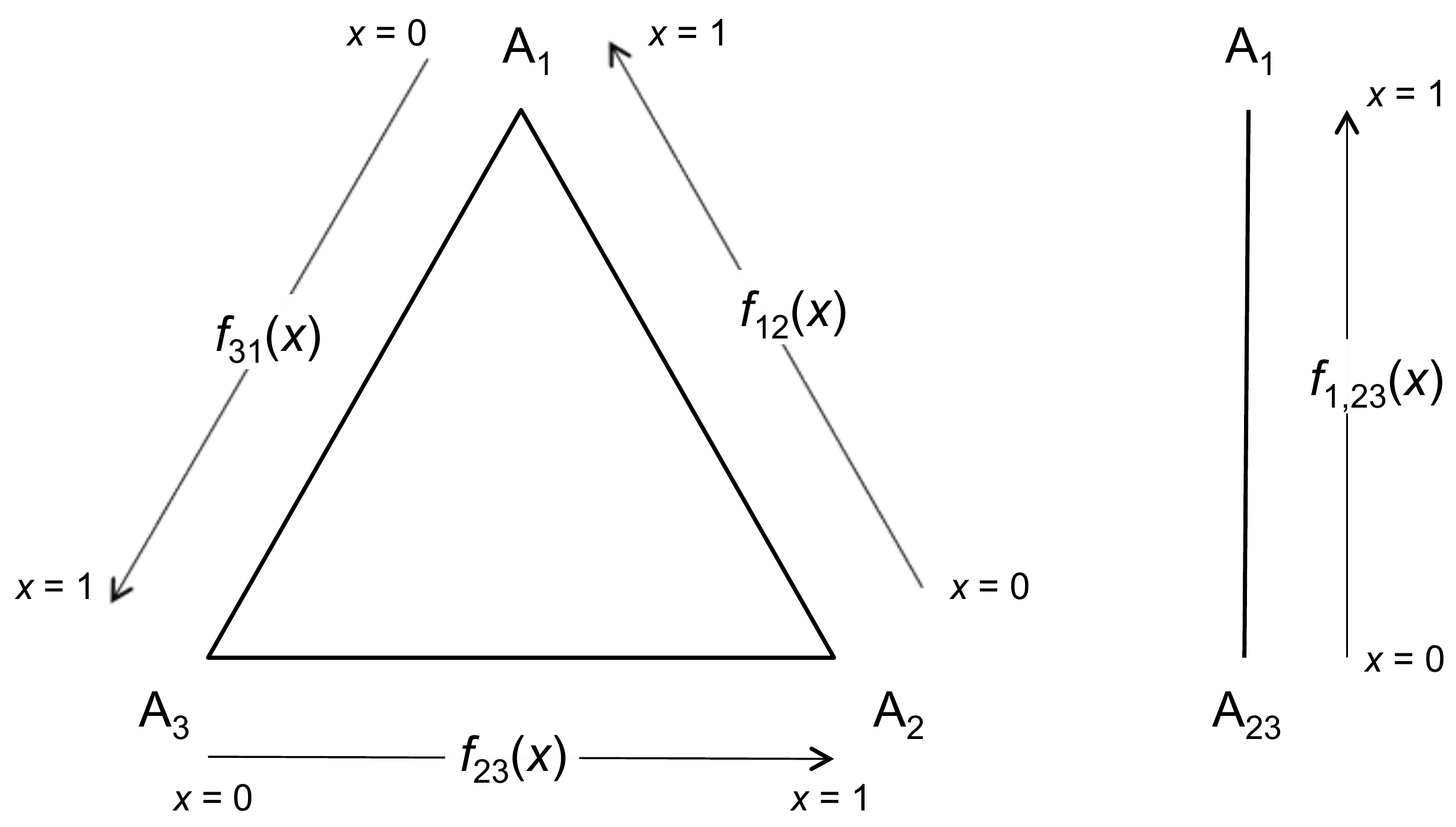} 
   \caption{The left-hand diagram is the 2-simplex over which the solution to the forward Kolmogorov equation for $K = 3$ alleles is defined.  
   The vertices labelled $A_1$, $A_2$ and $A_3$ 
   correspond to the states in which the entire population has allele $A_1$, $A_2$ or $A_3$ respectively at the genomic site in question.  
   The functions $f_{12}$, $f_{23}$ and $f_{31}$ are line densities approximating the stationary distribution to Eq.~(\ref{generalPDE})  
   on the edges indicated for a rate matrix with transition rates $<<1$.  
   The right-hand diagram is the 1-simplex supporting the effective 2-allele model corresponding to the allele partitioning $(1)(23)$.}
   \label{fig:2Simplex}
\end{figure}

Now assume that $\max(\alpha_{12}, \alpha_{13}, \alpha_{23}, \left| \Phi \right| ) << 1$, so that the off-diagonal elements of $Q$ are small and the solution 
$f(x_1, x_2)$ to Eq.~(\ref{generalPDE}) is concentrated on the boundary of the 2-simplex illustrated in Fig.~\ref{fig:2Simplex}.  
\textcolor{black}{
For $K = 3$, we demonstrate in \ref{sec:StationaryBoundary} that the stationary solution to the full forward Kolmogorov equation, Eq.~(\ref{generalPDE}), 
is well approximated by a line density of the form Eq.~(\ref{approx2AlleleSoln}) along each edge of the 2-simplex.  Thus we
}
define line densities $f_{12}(x)$, $f_{23}(x)$ and $f_{31}(x)$ of the site frequency spectrum as shown, 
where the argument $x$ refers to the proportion of $A_1$, $A_2$ and $A_3$-type alleles respectively at a given genomic site, and hence $(1 - x)$ is the 
proportion of $A_2$, $A_3$ and $A_1$-type alleles respectively.  

The form of the rate matrix $Q$ entails that there is a net flux $\Phi$ of probability circulating clockwise around the boundary of the simplex.  
Thus the line density along each edge takes the form of the 
\textcolor{black}{
approximate $K = 2$-allele solution Eq~(\ref{approx2AlleleSoln}),
} 
\begin{equation}
\begin{split}
f_{12}(x) = f(x; C_{12}, \Phi), \\
f_{23}(x) = f(x; C_{23}, \Phi), \\
f_{31}(x) = f(x; C_{31}, \Phi), 
\end{split}			\label{dEquals3Soln}
\end{equation}
where the normalisation constants $C_{12}$, $C_{23}$ and $C_{31}$ are yet to be determined.  

In order to determine the unknown constants we partition the three alleles into two subsets and relabel all alleles within a given subset as an effective single
allele as described in \ref{sec:Equiv2}.  
For instance, the partitioning $(1)(23)$ illustrated in Fig.~\ref{fig:2Simplex} yields a $K = 2$-state Markov model with transition matrix 
\begin{equation}
Q^{(1)(23)} = \left( 
\begin{array}{cc}
- Q_{1,23}	&	Q_{1,23}	\\
Q_{23,1}	&	-Q_{23,1}	
\end{array}
\right),								\label{equiv2By2QMatrix}
\end{equation}
where (see Eq.~(\ref{equiv2AlleleQ}))
\begin{equation}
Q_{1,23} = Q_{12} + Q_{13}, \qquad Q_{23,1} = \frac{\pi_2 Q_{21} + \pi_3 Q_{31}}{\pi_2 + \pi_3}. 				\label{equiv2By2Q}
\end{equation}
The stationary state of this matrix is $(\pi_1, \pi_2 + \pi_3)$.  
\textcolor{black}{
As pointed out in the discussion following Eq.~(\ref{qTildeDerivation}), an  effective 2-allele partitioning of the original 3-allele model is 
not Markovian and therefore does not have a time-dependent behaviour equivalent 
to that of the Markov chain defined by Eqs.~(\ref{equiv2By2QMatrix}) and (\ref{equiv2By2Q}).  However its stationary distribution {\em is} equivalent 
to that of the 2-allele Markov chain in the sense that transitions between effective states occur with the same frequency in both models once the 
stationary state is achieved.  
}

The effective 2-allele model leads to a stationary distribution  
\begin{equation}
f_{1,23}(x) = f(x; C_{1, 23}, 0) = C_{1,23} x^{2Q_{23,1} - 1} (1 - x)^{2Q_{1,23} - 1}	\label{effectiveSoln}
\end{equation}
where 
\begin{equation}
C_{1,23} = \frac{1}{B(2Q_{23,1}, 2Q_{1,23})}.  \label{C123Def}
\end{equation}
No $\Phi$ term is present because no flux can cross the boundary points at $x = 0, 1$ of the 2-allele model.  In fact this distribution depends only on 
$Q^{\rm GTR}$ and is independent of $Q^{\rm flux}$.  As suggested by Fig.~\ref{fig:2Simplex} we identify this 
density with the marginal distribution of the 3-allele site frequency spectrum, and thus 
\begin{equation}
f_{1,23}(x) \approx f_{12}(x) + f_{31}(1 - x),  \label{marginal}
\end{equation}
where we have indicated that the equality is not exact in the sense that the stationary distribution to Eq.~(\ref{generalPDE}) for $K = 3$ has been assumed to 
be concentrated on the boundary of the simplex.  The approximation is most accurate away from the corners of the simplex, which is also 
where the approximate solution Eq.~(\ref{approx2AlleleSoln}) is most accurate.  Applying Eq.~(\ref{approx2AlleleSoln}) and using Eqs.~(\ref{dEquals3Soln}) 
and (\ref{effectiveSoln}) one finds that both sides of Eq.~(\ref{marginal}) are, to a good approximation, proportional to $x^{-1}(1 - x)^{-1}$.  
Equating coefficients gives 
\begin{equation}
C_{1,23} = C_{12} + C_{31}.  \label{C1Comma23}
\end{equation}
Similarly the other two possible partitionings give 
\begin{equation}
C_{2,31} = C_{23} + C_{12}, \qquad C_{3,12} = C_{31} + C_{23}, \label{C231And312}
\end{equation}
where $C_{2,31}$ and $C_{3,12}$ are defined by cyclically permuting the indices in Eqs.~(\ref{C123Def}) and (\ref{equiv2By2Q}).  These equations 
solve to give the required normalisation constants as 
\begin{equation}
\begin{split}
C_{12} &= \tfrac{1}{2} (C_{1,23} + C_{2,31} - C_{3,12}) \\
C_{23} &= \tfrac{1}{2} (C_{2,31} + C_{3,12} - C_{1,23}) \\
C_{31} &= \tfrac{1}{2} (C_{3,12} + C_{1,23} - C_{2,31}). 			\label{CabSolution}
\end{split}
\end{equation}

For consistency, given that approximating the neutral Wright-Fisher stationary distribution as a set of line densities is intended to 
be a lowest order approximation in the off-diagonal elements of the rate matrix, one can further approximate Eq.~(\ref{C123Def}) and its 
cyclic permutations using the expansion
\begin{equation}
\frac{1}{B(\epsilon, \eta)} = \frac{\epsilon\eta}{\epsilon + \eta} \left(1 + O(\epsilon) + O(\eta)\right), \quad \epsilon, \eta \rightarrow 0\!+.  \label{invBetaApprox}
\end{equation}
Using Eqs.~(\ref{general3By3Q}), (\ref{equiv2By2Q}), (\ref{CabSolution}) and (\ref{invBetaApprox}), and the symbolic manipulation package Mathematica~\cite{Mathematica15} 
we obtain the lowest order approximate normalisations 
\begin{equation}
C_{ab} = 2 \alpha_{ab} \pi_a \pi_b,  \label{approxNorms}
\end{equation}
and thus
\begin{equation} 
f_{ab}(x) \approx \frac{2\alpha_{ab}\pi_a\pi_b}{x(1 - x)} - \Phi\left(\frac{1}{x} - \frac{1}{1 - x}\right). \label{lineDensityKEquals3}
\end{equation}
The restriction Eq.~(\ref{phiRestriction}) on $\Phi$ ensures that the approximate form of the line density on each edge is non-negative.  

\textcolor{black}{
To summarise, given any general rate matrix with off-diagonal elements $Q_{ab} << 1$ and stationary left eigenvector $\pi_a$, 
the stationary distribution to the neutral Wright-Fisher model can be represented as a line density 
\begin{equation}
f_{ab}(x) \approx (C_{ab} - \Phi)\frac{1}{x} + (C_{ab} + \Phi)\frac{1}{1 - x}, 	\label{altLineDensityKEquals3}
\end{equation} 
along each edge $a$-$b$, where $x$ is the relative proportion of $A_a$ alleles and 
\begin{equation} 
C_{ab} = \pi_a Q_{ab} + \pi_b Q_{ba}, \quad \Phi = \pi_a Q_{ab} - \pi_b Q_{ba}.  \label{CPhiDefKEquals3}
\end{equation}
The properties of $Q$ ensure that this final formula for $\Phi$ is independent of which edge $a$-$b$ is chosen.  
}

\begin{figure}[!]
   \centering
   \includegraphics[width=\textwidth]{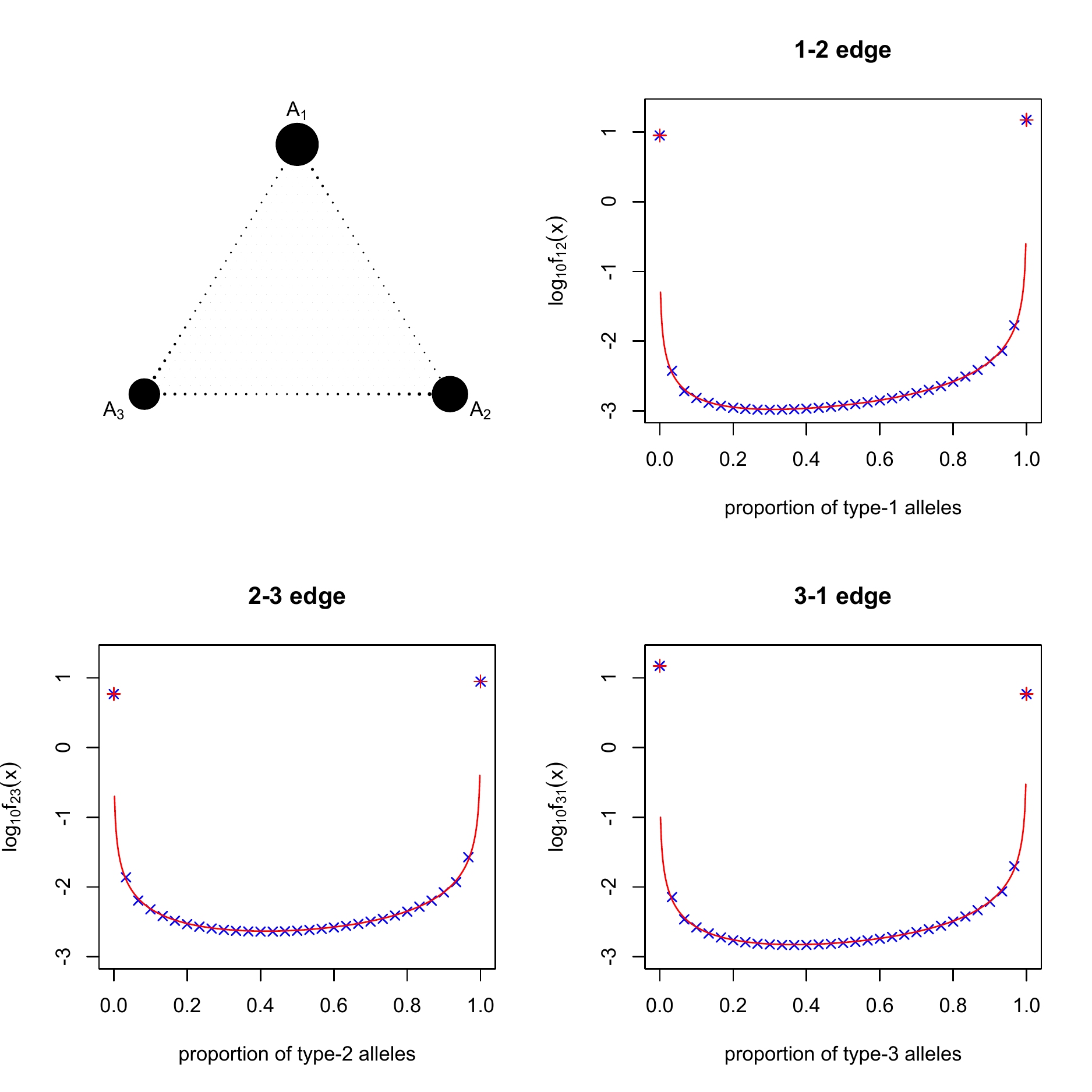} 
   \caption{Simulation of the neutral Wright-Fisher model with $K = 3$ alleles and a haploid population size $N = 30$.  
   The triangular pattern is the numerical determined stationary site frequency spectrum  
   of a 3-allele neutral Wright-Fisher with mutations.  Parameter values are $(\alpha_{12}, \alpha_{13}, \alpha_{23}) = (1, 2, 5)/1000$, 
   $(\pi_1, \pi_2, \pi_3) = (0.5, 0.3, 0.2)$ and $\phi= 0.2/1000$.    
   The full distribution over all $(N + 1)(N + 2)/2$ points is calculated, though points in the interior of the triangle are too small to be easily visible.  The remaining 
   three plots are the site frequency spectra on the boundary.  Blue crosses are the numerically determined stationary distribution multiplied by $N$.  The 
   red curves are the analytic solutions from Eq.~(\ref{lineDensityKEquals3}) and the red plus signs are $N$ times the theoretical probabilities 
   that a site is non-segregating, Eq.~(\ref{P_A1}) and (\ref{P_A2andA3}).}
   \label{fig:3allelesPlot}
\end{figure}

To test the accuracy of the analytic solution we have performed a simulation of the neutral Wright-Fisher model with $K = 3$ alleles and a full $3 \times 3$ 
mutation rate matrix.  In this simulation the population size is $N = 30$, and mutation rates from allele $a$ to allele $b$ for $a \ne b$ are $Q_{ab}/N$, 
where $Q_{ab}$ are elements of a rate matrix of the form Eq.~(\ref{general3By3Q}) with parameters $\alpha_{12} = 1/1000$, $\alpha_{13} = 2/1000$, 
$\alpha_{23} = 5/1000$, $(\pi_1, \pi_2, \pi_3) = (0.5, 0.3, 0.2)$ and $\Phi=0.2/1000$.  Results are shown in Fig.~\ref{fig:3allelesPlot}.  It is clear from 
the figure that the stationary distribution is almost completely concentrated on the boundary of the simplex and particularly heavily weighted at the corners, 
as expected.  The theoretical line densities $f_{12}$, $f_{23}$ and $f_{31}$ using the approximation Eq.~(\ref{lineDensityKEquals3}) 
are plotted together with the numerically determined stationary distribution, suitably 
normalised for comparison.  In general, by experimenting with a range of parameter values, we have found agreement between simulation and theory 
to be very good provided the elements of the rate matrix $Q$ are less than $10^{-2}$.  

Note that the line densities $f_{12}$, $f_{23}$ and $f_{31}$ are not suitable for evaluating the stationary distribution close to the corners of the simplex.  
The probability that a genomic site is a non-segregating site with allele of type $A_1$, say, can be calculated instead by integrating the effective 
2-allele distribution Eq.~(\ref{effectiveSoln}) from $1 - 1/N$ to $1$, where $N$ is the population size.  This gives 
\begin{eqnarray}
P(A_1) & = & \Prob(\mbox{site is non-segregating of type $A_1$}) \nonumber \\
	& = & \int_{1 - 1/N}^1 f_{1,23}(x) \, dx \nonumber \\
	& \approx & \frac{1}{2Q_{1,23} N^{2Q_{1,23}} B(2Q_{23,1}, 2Q_{1,23})} \nonumber \\
	& \approx & \pi_1 N^{-2(\alpha_{12}\pi_2 + \alpha_{13}\pi_3)},  \nonumber\\
	& \approx & \pi_1 \left\{ 1 - 2(\alpha_{12}\pi_2 + \alpha_{13}\pi_3)  \log N \right\}. 	\label{P_A1}
\end{eqnarray}
were we have used the approximation Eq.(\ref{invBetaApprox}) to the beta-function together with Eqs.~(\ref{general3By3Q}) and (\ref{equiv2By2Q}) 
in the second last line.  Similarly, the probability a site is non-segregating of type $A_2$ or $A_3$ is 
\begin{equation}
\begin{split}
P(A_2) \approx \pi_2 \left\{ 1 - 2(\alpha_{21}\pi_1 + \alpha_{23}\pi_3)  \log N \right\}, \\
P(A_3) \approx \pi_3 \left\{ 1 - 2(\alpha_{31}\pi_1 + \alpha_{32}\pi_2)  \log N \right\},  
\end{split} \label{P_A2andA3}
\end{equation}
respectively.  These probabilities are also plotted in Fig.~\ref{fig:3allelesPlot}, and agree very well with the numerical simulation.  

To complete the $K = 3$ case we check the normalisation of the approximate first order solution.  The total probability of the stationary distribution is  
\begin{equation}
\sum_{1 \le a < b \le 3} \int_{1/N}^{1 - 1/N} f_{ab}(x) \, dx + \sum_{a = 1}^3 P(A_a).  
\end{equation}
From Eq.~(\ref{lineDensityKEquals3}), the first term is 
\begin{equation}
\sum_{1 \le a < b \le 3} \int_{1/N}^{1 - 1/N} f_{ab}(x) \, dx
\approx 4 \sum_{1 \le a < b \le 3} \alpha_{ab} \pi_a\pi_b \log N,  
\end{equation} 
while the second term is, from Eqs.~(\ref{P_A1}) and(\ref{P_A2andA3}), 
\begin{equation}
\begin{split}
\sum_{a = 1}^3 P(A_a) 
& \approx \pi_1\left\{1 - 2(\alpha_{12}\pi_2 + \alpha_{13}\pi_3) \log N \right\} + \mbox{cyclic permutations} \\
& = 1 - 4 \sum_{1 \le a < b \le 3} \alpha_{ab} \pi_a\pi_b \log N. 
\end{split}
\end{equation}
The two terms sum to 1, as required.  


\section{$K = 4$ case}
\label{sec:K4Case}

\begin{figure}[t]
   \centering
   \includegraphics[width=0.8\textwidth]{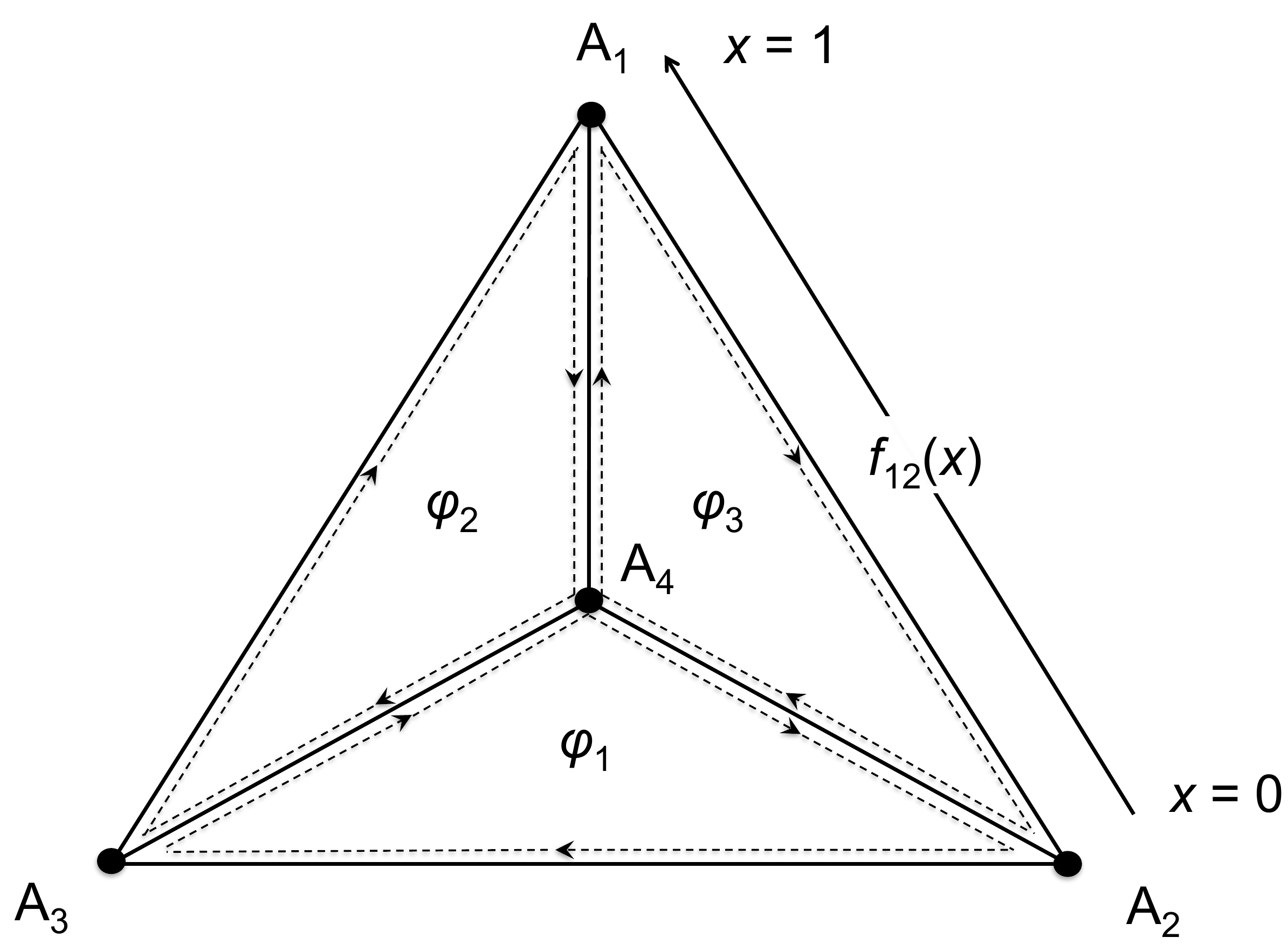} 
   \caption{The simplex over which the stationary solution to the $K = 4$ forward Kolmogorov equation is defined.  The solution is represented as a set of 
   line densities $f_{ab}(x)$ along the edges, with $f_{12}(x)$ shown as an example.   
   $\Phi_1$, $\Phi_2$ and $\Phi_3$ are the net probability fluxes around three triangular paths shown.}
   \label{fig:Simplex4Alleles}
\end{figure}

The $K = 4$ case is relevant to the DNA alphabet $\{A, C, G, T\}$ and therefore to analysis of genomic site frequency spectra.  
As for the $K = 3$ case, we consider the most general rate matrix, the only restrictions being that off diagonal elements are non-negative and rows sum to zero.
Any such rate matrix is specified by 12 independent parameters can be split into a reversible part and a `flux' part:  
\begin{equation}
Q = Q^{\rm GTR} + Q^{\rm flux}	.	\label{generalQ}
\end{equation}
The reversible part has 9 independent parameters and can be written as~\cite{tavare1986some} 
\begin{equation}
Q^{\rm GTR} = \left(
\begin{array}{cccc}
\bullet			& \alpha_{12}\pi_2	& \alpha_{13}\pi_3	& \alpha_{14}\pi_4	\\
\alpha_{21}\pi_1	& \bullet			& \alpha_{23}\pi_3	& \alpha_{24}\pi_4	\\
\alpha_{31}\pi_1	& \alpha_{32}\pi_2	& \bullet			& \alpha_{34}\pi_4	\\
\alpha_{41}\pi_1	& \alpha_{42}\pi_2	& \alpha_{43}\pi_3	& \bullet	
\end{array}
\right) 		\label{QGTRForKEquals4}
\end{equation}
where $\alpha_{ab} = \alpha_{ba}$ for $a, b = 1, \ldots, 4$, $a \ne b$, and the diagonal elements are set by ensuring the rows sum to $0$.  The row vector 
$(\pi_1, \pi_2, \pi_3, \pi_4)$ is the stationary distribution of the continuous time Markov model, normalised so that $\sum_{i = 1}^4 \pi_i = 1$.  The flux part has 3 
parameters and takes the form: 
\begin{equation}
Q^{\rm flux} = \frac{1}{2}\sum_{i = 1}^3 \Phi_i F^{(i)} 
\end{equation}
where 
\begin{gather}
F^{(1)} = \left(
\begin{array}{cccc}
0	&	0		&	0			&	0		\\
0	&	0		&	 1/\pi_2		&	-1/\pi_2	\\
0	&	-1/\pi_3	&	0			&	 1/\pi_3	\\
0	&	 1/\pi_4	&	-1/\pi_4		&	0	
\end{array}
\right), \quad
F^{(2)} = \left(
\begin{array}{cccc}
0		&	0	&	-1/\pi_1	&	 1/\pi_1	\\
0		&	0	&	0		&	0		\\
 1/\pi_3	&	0	&	0		&	-1/\pi_3	\\
-1/\pi_4	&	0	&	 1/\pi_4	&	0	
\end{array}
\right), \nonumber \\  \\
F^{(3)} = \left(
\begin{array}{cccc}
0		&	 1/\pi_1	&	0	&	-1/\pi_1	\\
-1/\pi_2	&	0		&	0	&	 1/\pi_2	\\
0		&	0		&	0	&	0		\\
 1/\pi_4	&	-1/\pi_4	&	0	&	0	
\end{array}
\right). \nonumber
\end{gather}
The parameters $\Phi_i$ represent a net rate of transitions around each of three independent triangular paths along edges of  the 4-allele simplex 
over which the solution of the forward Kolmogorov equation is defined (see Fig.~\ref{fig:Simplex4Alleles}).  The path around the fourth triangular face, 
namely $A_1 \rightarrow A_2 \rightarrow A_3 \rightarrow A_1$, can be written as a sum of paths around the other three faces.   
The requirement that the off-diagonal elements of $Q$ be positive implies that the following constraints must also hold:
\begin{equation}
\begin{split}
&  \left| \Phi_1 \right| \le 2\alpha_{23}\pi_2\pi_3 \\
&  \left| \Phi_2 \right| \le 2\alpha_{13}\pi_1\pi_3 \\
&  \left| \Phi_3 \right| \le 2\alpha_{12}\pi_1\pi_2 \\
&  \left| \Phi_1- \Phi_2\right| \le 2\alpha_{34}\pi_3\pi_4 \\
&  \left| \Phi_3- \Phi_1\right| \le 2\alpha_{24}\pi_2\pi_4 \\
&  \left| \Phi_2- \Phi_3\right| \le 2\alpha_{14}\pi_1\pi_4 \\
\end{split}
\end{equation}

\textcolor{black}{
Under the assumption that $\max(\alpha_{ab}, \left|\Phi_a\right|) << 1$, we can again approximate solution of the forward Kolmogorov equation as 
a set of line densities on the edges of the simplex defined by Eq.~(\ref{simplex}), where the line density on each edge has the form of the 
function $f(x; C_{ab}, \Phi_{ab})$ defined by Eq.~(\ref{approx2AlleleSoln}).  Consider, for instance, the $A_1$-$A_2$ edge of the simplex.  
By first partitioning the alleles into 3 effective alleles $A_1$, $A_2$, and $A_{(3,4)} = \{A_3, A_4\}$, we can first construct an effective 3-allele model 
using arguments analogous to those in \ref{sec:Equiv2}.  Then, following the logic of \ref{sec:StationaryBoundary} we see that the effective 3-allele model 
has a stationary distribution which can be approximated along the $A_1$-$A_2$ boundary by a line density of the appropriate form.  
}

Thus we have a set of six functions 
\begin{equation}
f_{ab}(x) = f(x; C_{ab}, \Phi_{ab}), \qquad  1 \le a < b \le 4
\end{equation}
defined using the convention that along the edge $(ab)$, $x$ is the relative proportion of type-$a$ alleles and $1 - x$ is the relative proportion of type-$b$ 
alleles, as illustrated in Fig.~\ref{fig:Simplex4Alleles} for the edge $(ab) = (12)$.  From the diagram we read off the flux parameters 
\begin{equation}
\begin{split}
\Phi_{12} &= \Phi_3 \\
\Phi_{13} &= -\Phi_2 \\
\Phi_{14} &= \Phi_2 - \Phi_3 \\
\Phi_{23} &= \Phi_1 \\
\Phi_{24} &= \Phi_3 - \Phi_1 \\
\Phi_{34} &= \Phi_1 - \Phi_2 	\label{phiab}
\end{split}
\end{equation}

Following the procedure used for the $K = 3$ case, the normalisations $C_{ab}$ are determined by partitioning the alleles into two distinct subsets and 
equating the corresponding marginal distributions with the stationary distribution of the equivalent 2-allele model, as defined in \ref{sec:Equiv2}.  
There are seven possible partitionings, leading to the seven equations 
\begin{equation}
\begin{split}
C_{1,234} &= C_{12} + C_{13} + C_{14} \\
C_{2,341} &= C_{23} + C_{24} + C_{12} \\
C_{3,412} &= C_{34} + C_{13} + C_{23} \\
C_{4,123} &= C_{14} + C_{24} + C_{34} \\
C_{12,34} &= C_{13} + C_{14} + C_{23} + C_{24} \\
C_{13,24} &= C_{12} + C_{14} + C_{23} + C_{34} \\
C_{14,23} &= C_{12} + C_{13} + C_{24} + C_{34},		\label{CNormEquations}
\end{split}
\end{equation}
where 
\begin{equation}
\begin{split}
C_{a,bcd} = \frac{1}{B(2Q_{bcd,a}, 2Q_{a,bcd})} \approx \frac{2Q_{bcd,a} Q_{a,bcd}}{ Q_{bcd,a} + Q_{a,bcd}}, \\
C_{ab,cd} = \frac{1}{B(2Q_{cd,ab}, 2Q_{ab,cd})} \approx \frac{2Q_{cd,ab} Q_{ab,cd}}{ Q_{cd,ab} + Q_{ab,cd}}. 			\label{C4IndexDef}
\end{split}
\end{equation}
The elements of the effective $2 \times 2$ rate matrix are, from Eq.~(\ref{equiv2AlleleQ}), 
\begin{equation}
\begin{split}
Q_{a, bcd} &= Q_{ab} + Q_{ac} + Q_{ad}\\
Q_{bcd, a} &= \frac{\pi_{b}Q_{ba} + \pi_{c}Q_{ca} + \pi_{d}Q_{da}}{\pi_{b} + \pi_{c} + \pi_{d}}\\
Q_{ab, cd} &=  \frac{\pi_{a}(Q_{ac} + Q_{ad}) + \pi_{b}(Q_{bc} + Q_{bd})}{\pi_{a} + \pi_{b}}.
\end{split}
\end{equation}
At first sight the system of equations (\ref{CNormEquations}) appears to be overdetermined.  However, we have determined using 
Mathematica~\cite{Mathematica15} that provided one uses the first order approximation to the beta-function in Eq.~(\ref{C4IndexDef}), the equations are not 
independent, but yield the solution  
\begin{equation}
C_{ab} = 2\alpha_{ab} \pi_a \pi_b,      \label{lineDensityNormalisations}
\end{equation}
irrespective of which subset of six equations is used.  Note that the $C_{ab}$ depend only on parameters defining $Q^{\rm GTR}$, and not on $Q^{\rm flux}$.  

The solutions $P(A_a)$ at the corners of the simplex corresponding to the probabilities that a site is non-segregating and of allele type $A_a$ in a population 
of size $N$ are found by analogy with Eq.~(\ref{P_A1}) to be 
\begin{eqnarray}
P(A_a) & \approx & \pi_a N^{-2\sum_{b\ne a}\alpha_{ab}\pi_b} \nonumber \\
              & \approx & \pi_a \left( 1 - 2\sum_{b\ne a}\alpha_{ab}\pi_b \log N \right).  \label{cornerProbabilities}
\end{eqnarray}

\begin{figure}[!]
   \centering
   \includegraphics[width=\textwidth]{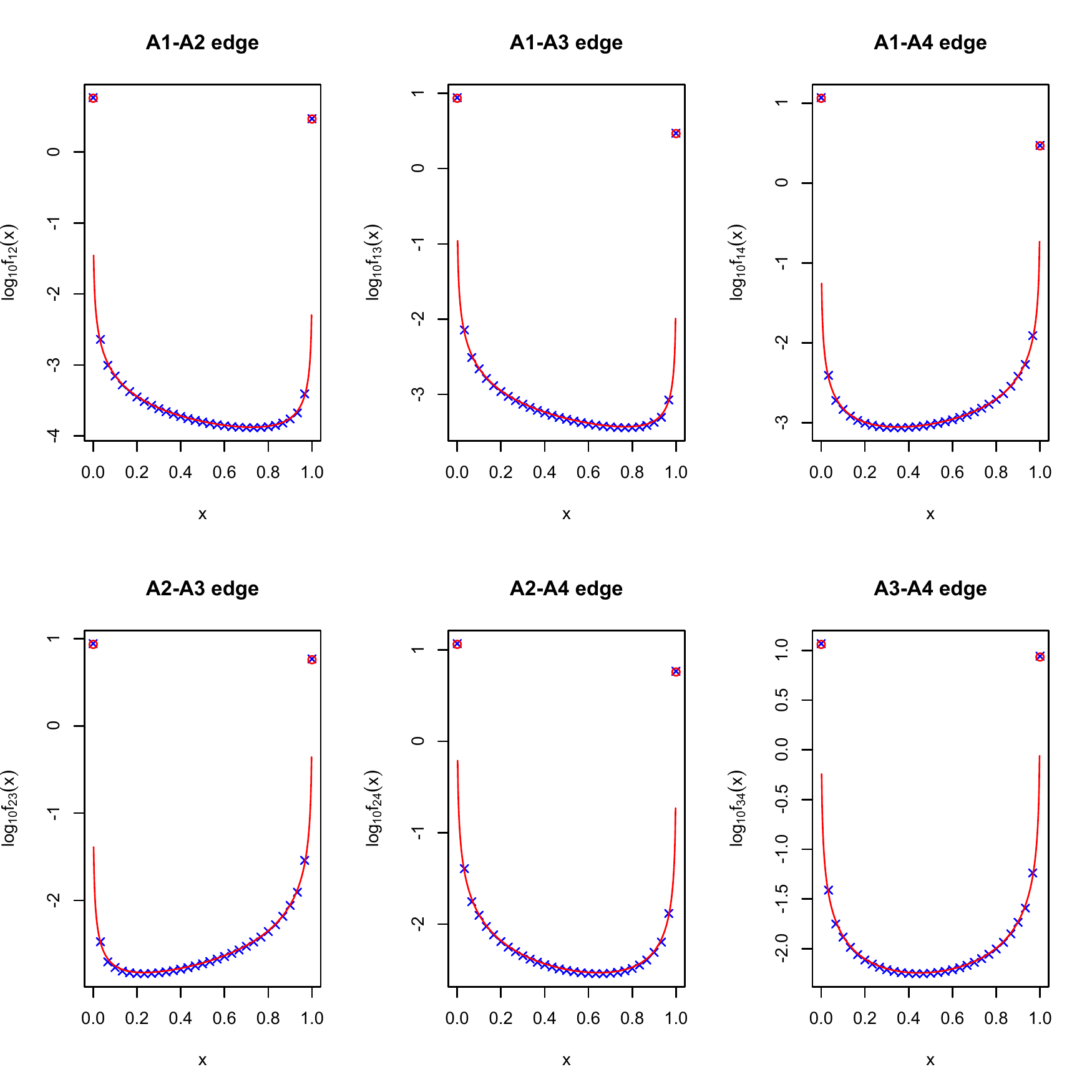} 
   \caption{Simulation of the neutral Wright-Fisher model with $K = 4$ alleles and a population size $N = 30$. 
   Blue crosses are the numerically determined stationary distribution along 
   each edge of the simplex over which the site frequency distribution is defined. For any 2 alleles, $A_a$ and $A_b$, the parameter $x$ is the relative proportion 
   of $A_a$ alleles and $1 - x$ is the relative proportion of $A_b$ alleles.   Superimposed in red are the theoretical line densities $f_{ab}(x)$, 
   Eq.~(\ref{approx2AlleleSoln}) with coefficients on the edge $(ab)$ given by Eqs.~(\ref{lineDensityNormalisations}) and (\ref{phiab}), 
   plotted on a logarithmic scale.  The red circles are 
   the theoretical probabilities at the simplex corners,  Eq.~(\ref{cornerProbabilities}). Parameter values of the rate matrix $Q$ are 
   $(\alpha_{12}, \alpha_{13}, \alpha_{14}, \alpha_{23}, \alpha_{24}, \alpha_{34}) = 0.001 \times (1, 2, 3, 4, 5, 6)$,  $(\pi_1, \pi_2, \pi_3, \pi_4) = (0.1, 0.2, 0.3, 0.4)$ 
   and $(\Phi_1, \Phi_2, \Phi_3) = 0.001 \times (0.4, 0.1, -0.03)$.}
   \label{fig:4allelesPlot1}
\end{figure}

\begin{figure}[!]
   \centering
   \includegraphics[width=\textwidth]{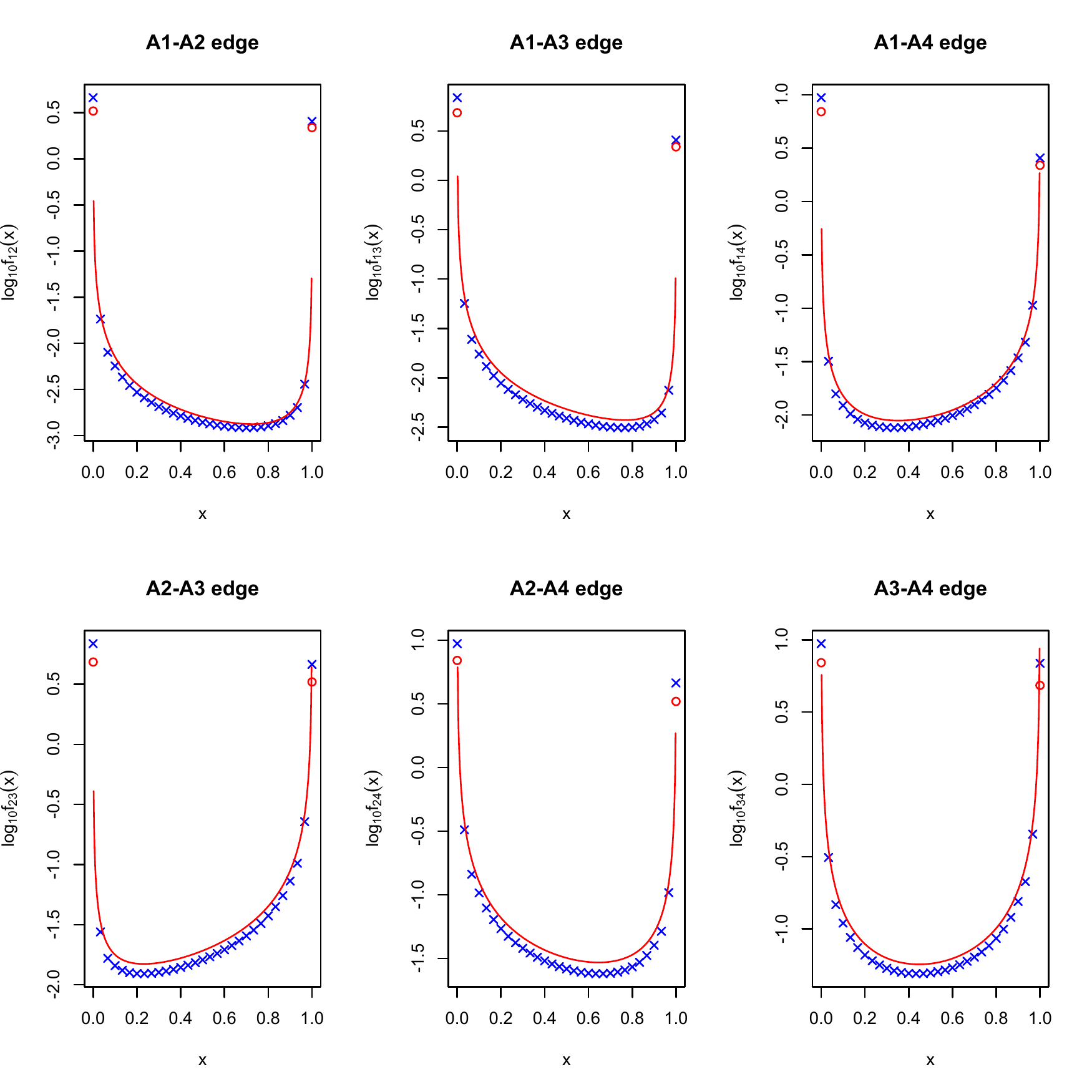} 
   \caption{The same as Fig.~\ref{fig:4allelesPlot1}, but with rate matrix parameters 
   $(\alpha_{12}, \alpha_{13}, \alpha_{14}, \alpha_{23}, \alpha_{24}, \alpha_{34}) = 0.01 \times (1, 2, 3, 4, 5, 6)$,  $(\pi_1, \pi_2, \pi_3, \pi_4) = (0.1, 0.2, 0.3, 0.4)$ 
   and $(\Phi_1, \Phi_2, \Phi_3) = 0.01 \times (0.4, 0.1, -0.03)$}
   \label{fig:4allelesPlot2}
\end{figure}

The accuracy of the approximate solution is illustrated in Figs.~\ref{fig:4allelesPlot1} and \ref{fig:4allelesPlot2}.  These plots show simulations of the 
stationary distribution of the 4-allele neutral Wright-Fisher model defined by Eq.~(\ref{eq:FullPij}) for a population of $N = 30$.  The mutation rates $u_{ab}$ 
in Eqs.~(\ref{psiDef}) and (\ref{QToU}) correspond to an instantaneous rate matrix $Q$ with parameters 
$(\alpha_{12}, \alpha_{13}, \alpha_{14}, \alpha_{23}, \alpha_{24}, \alpha_{34}) = (1, 2, 3, 4, 5, 6) \times \theta$,  $(\pi_1, \pi_2, \pi_3, \pi_4) = (0.1, 0.2, 0.3, 0.4)$ 
and $(\Phi_1, \Phi_2, \Phi_3) = (0.4, 0.1, -0.03) \times \theta$ where $\theta = 0.001$ in Fig.~\ref{fig:4allelesPlot1} and $0.01$ in Fig.~\ref{fig:4allelesPlot2}.  
Superimposed in red are the approximate theoretical line densities, Eq.~(\ref{approx2AlleleSoln}) with coefficients on the edge $(ab)$ given by Eqs.~(\ref{lineDensityNormalisations}) and (\ref{phiab}), 
and the probabilities at the simplex corners,  Eq.~(\ref{cornerProbabilities}).  As a rule of thumb we find that for these parameters and for a range of other 
parameter values that we have tried, the agreement between simulation and theory is very  close provided the off-diagonal elements of $Q$ are less than 
$10^{-2}$, as in Fig.~\ref{fig:4allelesPlot1}, but fails to be close for higher mutation rates, as in Fig.~\ref{fig:4allelesPlot2}.  
\textcolor{black}{
More specifically, in Fig.~\ref{fig:4allelesPlot2} one sees that the theoretical solution under-estimates slightly at the corners as it fails to account for 
non-zero probability in the interior of the simplex near each corner, corresponding to rare 3- or 4-allele SNPs.  The normalisation of the complete distribution to 
unity then causes the approximate line densities to be correspondingly overestimated.
}

\section{Arbitrary $K$}
\label{sec:ArbitraryK}

The analogous procedure can in principle be followed for a general $K \times K$ rate matrix $Q$ with $K(K - 1)$ parameters for any value of $K \ge 3$ by 
parameterising  $Q$  as a sum of a reversible part and a flux part, as in Eq.~(\ref{generalQ}).  For the reversible part an appropriate parameterisation for the elements 
of $Q^{\rm GTR}$ is~\cite{tavare1986some}
\begin{equation}
Q_{ab}^{\rm GTR} = \begin{cases}
\alpha_{ab} \pi_b, & \mbox{if } a\ne b, \\
-\sum_{c \ne a}  \alpha_{ac} \pi_c & \mbox{if } a = b, 
\end{cases}
\end{equation}
for parameters $\pi_b$ and $\alpha_{ab}$, $a \ne b = 1, \ldots K$, subject to the constraints $\alpha_{ab} = \alpha_{ba} \ge 0$, $\pi _b \ge 0$, and 
$\sum_{b = 1}^K \pi_b = 1$.   The number of independent parameters required to define $Q^{\rm GTR}$ is 
\begin{equation}
\frac{K(K - 1)}{2} + (K - 1) = \frac{(K - 1)(K + 2)}{2}.	\label{NumberParametersGTR}
\end{equation}

For the flux part, an appropriate parameterisation is 
\begin{equation}
Q^{\rm flux} = \frac{1}{2}\sum_{1 \le i < j \le K - 1} \Phi_{ij} F^{(ij)}, 
\end{equation}
where the $F^{(ij)}$ are a set of $K \times K$ matrices whose elements are 
\begin{equation}
F_{ab}^{(ij)} = 
\{(\delta_{ia}\delta_{jb} - \delta_{ib}\delta_{ja}) 	
+ \delta_{aK}(\delta_{ib} - \delta_{jb}) + \delta_{bK}(\delta_{ja} - \delta_{ia})\} /\pi_a ,
\end{equation}
for $1 \le i < j \le K - 1$ and $a, b = 1, \ldots K$.  In the limit of small mutation rates $\Phi_{ij}$ represents a net flux of probability around the closed path 
$K \rightarrow i \rightarrow j \rightarrow K$ along the edges of the simplex over which the solution to Eq.~(\ref{eq:FullPij}) is defined.  
A flux around any other closed path $i \rightarrow j \rightarrow k \rightarrow i$ of length 3 can be written as a sum of these fluxes.  The number of independent 
fluxes is $(K - 1)(K - 2)/2$, which, together with Eq.~(\ref{NumberParametersGTR}) implies the total number of parameters defining $Q$ is  
$$
\frac{(K - 1)(K + 2)}{2} + \frac{(K - 1)(K - 2)}{2} = K(K - 1), 
$$
as required for a general rate matrix.  
\textcolor{black}{
It is easy to check that $\pi^{\rm T} Q^{\rm GTR} = \pi^{\rm T} Q^{\rm flux} = 0$ and hence that 
 that $(\pi_1 \ldots \pi_K)$ is the stationary distribution of $Q$.  
 }

The notation for the flux part of the rate matrix can be reconciled with the notation used in Section~\ref{sec:K4Case} for the $K = 4$ case 
using the relationships  
\begin{equation}
F^{(ij)} = \sum_{k = 1}^3 \epsilon_{ijk} F^{(k)}, \qquad \Phi_{ij} = \sum_{k = 1}^3 \epsilon_{ijk} \Phi_k, \label{epsIJK}
\end{equation}
where $\epsilon_{ijk}$ is the Levi-Civita symbol.  

In the limit that the off-diagonal elements of $Q$ are $<< 1$, the approximate stationary distribution of the neutral Wright-Fisher model is given as a line 
density $f_{ab}(x)$ along each edge $(ab)$ of the simplex as 
\begin{equation}
f_{ab}(x) = (C_{ab} - \Phi_{ab})\frac{1}{x} + (C_{ab} + \Phi_{ab})\frac{1}{1 - x}, 	\label{lineDensityArbitraryK}
\end{equation}
where $x$ is the relative proportion of $A_a$ alleles, $1 - x$ is the relative proportion of $A_b$ alleles, and 
\begin{equation}
\begin{split}
C_{ab} &= 2\alpha_{ab}\pi_a\pi_b = \pi_a Q_{ab} + \pi_b Q_{ba}, \\ 
\Phi_{ab} &= 2\pi_a Q_{ab}^{\rm flux} = \pi_a Q_{ab} - \pi_b Q_{ba},  \label{CPhiGeneralK}
\end{split}
\end{equation}
where $Q_{ab}^{\rm flux}$ are the elements of the matrix $Q^{\rm flux}$.  A necessary requirement for the approximate line density to be accurate is that 
\begin{equation}
\max_{a\ne b} Q_{ab} \times \left|\log x + \log(1 - x)\right| << 1,
\end{equation}  
where $Q_{ab}$ are the elements of the rate matrix $Q$.   
At the corners of the simplex, the probability that a site is non-segregating and 
of allele type $A_a$ in a population of size $N$ is  
\begin{equation}
P(A_a) \approx \pi_a \left( 1 - 2\sum_{b\ne a}\alpha_{ab}\pi_b \log N \right) = \pi_a  - \sum_{b\ne a}C_{ab} \log N.  \label{cornersArbitraryK}
\end{equation}


\section{\textcolor{black}{Strand Symmetry}}
\label{sec:StrandSymmetry}

Most genomic sequences, when examined on a sufficiently large scale, are observed to be strand symmetric, that is, symmetric under simultaneous 
interchange of nucleotides $A$ with $T$ and $C$ with $G$.  This symmetry appears to result 
from a spectrum of causes, not only mutation rates~\cite{baisnee2002complementary}.  Nevertheless it is interesting to explore the effect of a 
strand symmetric mutation rate matrix on the symmetries of the neutral-evolution site frequency spectrum.  

The most general strand-symmetric rate matrix has 6 independent parameters and takes a form in which the third and fourth rows are permutatons of the 
second and first rows respectively: 

\begin{equation}
Q^{\rm SS} = \left( \begin{array}{cccc}
\bullet	& Q_{AC}	& Q_{AG}	& Q_{AT}	\\
Q_{CA}	& \bullet	& Q_{CG}	& Q_{CT}	\\
Q_{CT}	& Q_{CG}	& \bullet	& Q_{CA}	\\
Q_{AT}	& Q_{AG}	& Q_{AC}	& \bullet	
\end{array}
\right) ,		\label{QSS}
\end{equation}
where the each diagonal element is minus the sum of the other three elements in its row.  It is easy to check that stationary distribution of this matrix is 
\begin{equation}
\pi^{\rm T} = \tfrac{1}{2}(\eta, 1 - \eta, 1 - \eta, \eta), 
\end{equation}
where 
\begin{equation}
\eta = \frac{Q_{CA} + Q_{CT}}{Q_{CA} + Q_{CT}+ Q_{AC} + Q_{AG}}.  
\end{equation}
For the three fluxes $\Phi_{AC}$, $\Phi_{CG}$ and $\Phi_{GA}$ defined by Eq.~(\ref{epsIJK}), substitution into Eq.~(\ref{CPhiGeneralK}) gives  
\begin{equation}
\Phi_{AC} = \Phi_{GA} = \frac{Q_{CT} Q_{AC} - Q_{AG} Q_{CA}}{2(Q_{CA} + Q_{CT}+ Q_{AC} + Q_{AG})}, \qquad \Phi_{CG} = 0.  
\end{equation}

Thus a non-reversible strand-symmetric rate matrix has only one independent flux corresponding to the closed path 
$A \rightarrow C \rightarrow T \rightarrow G \rightarrow A$ along the edges of the tetrahedron in Fig.~\ref{fig:Tetrahedron}.  It follows that, 
in a neutrally-evolving strand-symmetric Wright-Fisher model, the site-frequency spectrum is expected to be symmetric along the $AT$ and $CG$ 
edges, but may be asymmetric along the remaining edges if the rate matrix is non-reversible.  


\section{Discussion and Conclusions}
\label{sec:Conclusions}

We have obtained approximate solutions for the diffusion limit of the stationary distribution of the $K$-allele neutral Wright-Fisher model in the realistic limit 
of small scaled mutation rates for arbitrary values of $K$.  The solution is obtained in terms of a parameterisation in which the rate matrix $Q$ 
is decomposed into the sum of a general time-reversible part $Q^{\rm GTR}$ and a non-reversible `flux' part $Q^{\rm flux}$.   
The solutions consist of the line densities Eq.~(\ref{lineDensityArbitraryK}) defined on the 
edges of the simplex over which the stationary distribution is defined,  Eq.~(\ref{simplex}).  
The approximate line densities lose accuracy, and are not integrable, as the corners of the solution are approached.  
To overcome this, cutoffs can be imposed at both ends of the edge so that the density is defined on the interval $[1/N, 1 - 1/N]$, where $N$ is an effective 
population size, and the solution can be represented as the point masses Eq.~(\ref{cornersArbitraryK}) at the corners of the simplex.  

The ultimate aim of this work is to estimate rate matrices from allele frequency spectra obtained by genotyping moderate sized samples of a large population.  
\textcolor{black}{
Similar estimates of mutation and selection rate parameters have been obtained for an effective 2-allele model by 
Vogl and Bergman~\cite{vogl2015inference} using as little as 10 haploid whole genome {\em Drosophila simulans} sequences.
}
Note that in any genotyping data set the sample size is not the population size parameter $N$ used in the above calculations.  
The parameter $N$ should be regarded as essentially infinite in the diffusion limit despite the fact that the high dimensionality of the state space has 
restricted our numerical simulations to small population sizes, and that, nevertheless, the diffusion limit was observed to be approached very rapidly 
in $N$.  
\textcolor{black}{
(For instance, Figs.~\ref{fig:3allelesPlot} and \ref{fig:4allelesPlot1} show that the finite population simulation for $N = 30$ is almost indistinguishable from the 
diffusion limit.)  
}

In the case of $K = 2$ alleles, Vogl~\cite{vogl2014estimating} has solved the problem of estimating both parameters of the $2 \times 2$ rate matrix 
from an empirical allele frequency spectrum obtained from a finite sample of $M$ haplotypes, where $M << N$.  For higher values of $K$, his solution can be applied directly to each 
member of the set of effective 2-allele models defined by partitioning the alleles into distinct complementary subsets as described in \ref{sec:Equiv2}.  
\textcolor{black}{
This procedure is followed in~\cite{vogl2015inference} for the 4-letter genomic alphabet and the partitioning $(AT)(CG)$.  
}
From these estimates of effective $2 \times 2$ rate-matrix parameters an estimate of the reversible part $Q^{\rm GTR}$ of the full $K \times K$ can be constructed 
as follows.  Consider for instance the $K= 4$ case, in which the partitionings are of the form $\{a, b, c\} \cup \{d\}$ or $\{a, b\} \cup \{c, d\}$.  In Section~4.1 
of~\cite{vogl2014estimating} Vogl provides unbiassed maximum-likelihood estimates of the two independent parameters defining a $2 \times 2$ rate matrix. These 
estimates are given in terms of the number of sampled haplotypes $M$ and the relative proportions of 
segregating and non-segregating sites.  The first estimated parameter, labelled $\vartheta$, 
is the equivalent in our notation to the combinations occurring on the right-hand sides of of Eq.~(\ref{C4IndexDef}), which, for slow mutation rates, are the 
normalisations $C_{a,bcd}$ or $C_{ab,cd}$.  From these, one can use Eqs.~(\ref{CNormEquations}) and (\ref{lineDensityNormalisations}) to obtain 
unbiassed estimates of the combinations $2\alpha_{ab}\pi_a\pi_b$.   The second parameter defined by Vogl, referred to as the `mutation bias', 
translates to $\pi_d$, or $\pi_c  + \pi_d$ in our notation, depending on the partitioning.  From Eq.~(\ref{QGTRForKEquals4}) we therefore have an estimate of all 
parameters occurring in $Q^{\rm GTR}$.  

The non-reversible part $Q^{\rm flux}$ cannot be estimated from the above method, or, for that matter, from simple proportions of 
segregating and non-segregating sites.  From the line-densities plotted in Figs.~\ref{fig:3allelesPlot}, \ref{fig:4allelesPlot1} and \ref{fig:4allelesPlot2} 
it is clear that information about $Q^{\rm flux}$ is contained instead in the asymmetries of the spectrum across the intermediate range of frequencies 
along each edge of the simplex on which the allele-frequency spectrum is defined.  Development of estimators of the parameters of $Q^{\rm flux}$ 
following this line of attack is the subject of the authors' ongoing work.


\section*{Software}

The R programs used to produce Figures~\ref{fig:3allelesPlot}, \ref{fig:4allelesPlot1} and \ref{fig:4allelesPlot2} 
\textcolor{black}{
and the Mathematica programs used to 
derive Eqs.~(\ref{approxNorms}) and (\ref{lineDensityNormalisations})  
} 
are available at \\{\tt https://github.com/cjb105/SimulateNeutralWF}.  


\section*{Acknowledgments}

The authors wish to thank Claus Vogl and Jurag Bergman for encouraging us to investigate the asymptotic behaviour the forward Kolmogorov 
equation near the simplex boundary.  The results of this investigation are reported in \ref{sec:StationaryBoundary}.   We also wish to thank 
Asger Hobolth for a very thorough proof-read of the original manuscript. 


\appendix

\section{
\textcolor{black}{
Stationary solution to the $K = 3$ forward Kolmogorov equation near the simplex boundary}
\label{sec:StationaryBoundary}
}

In this Appendix we consider the asymptotic behaviour of the stationary solution to the full forward Kolmogorov equation in the vicinity of an edge of the 
simplex in Fig.~\ref{fig:2Simplex}, but not immediately close to the corners.  For $K = 3$ the stationary solution satisfies, from Eq.~(\ref{generalPDE}),  
\begin{equation}
\begin{split}
0  = 	 & \,\frac{1}{2} \frac{\partial^2}{\partial {x_1}^2} \left[ x_1 (1 - x_1) f \right] - \frac{\partial^2}{\partial {x_1}\partial {x_2}}(x_1 x_2 f) 
	+    \frac{1}{2} \frac{\partial^2}{\partial {x_2}^2} \left[ x_2 (1 - x_2) f \right]  \\
	& - \, \frac{\partial}{\partial x_1} \left( \sum_{b = 1}^3 x_b Q_{b1}f \right) - \frac{\partial}{\partial x_2} \left( \sum_{b = 1}^3 x_b Q_{b2}f \right). 
																			\label{3dKolmogorovEqn}
\end{split}
\end{equation}

Since we are interested in slow mutation rates, we introduce an overall mutation rate $\theta << 1$ defined by 
\begin{equation}
Q_{ab} = \theta q_{ab},\quad \sum_{a \ne b} q_{ab} = 1.
\end{equation}
Without loss of generality, consider the $A_1$-$A_2$ edge of the simplex.  Define new coordinates $(x, y)$ via 
\begin{equation}
(x_1, x_2, x_3) = (x - \tfrac{1}{2} y, 1 - x - \tfrac{1}{2} y, y).   
\end{equation} 
We are interested in determining the form of the stationary solution in a region 
\begin{equation}
\left| \log x + \log(1 - x) \right| << \theta^{-1} , \qquad 0 < y \le \Lambda, 	\label{XYRegion}
\end{equation}
which is close to the $A_1$-$A_2$ edge, but sufficiently far from the corners that we can expect to be able to recover the line density Eq.(\ref{approx2AlleleSoln}).  
The cutoff $\Lambda$ is not necessarily small.  In the new coordinates the gradient operators are 
\begin{equation}
\frac{\partial}{\partial x_1} = \frac{1}{2} \frac{\partial}{\partial x} - \frac{\partial}{\partial y}, \qquad 
\frac{\partial}{\partial x_2} = - \,\frac{1}{2} \frac{\partial}{\partial x} - \frac{\partial}{\partial y}. 
\end{equation}
A straightforward but lengthy calculation gives Eq.~(\ref{3dKolmogorovEqn}) in terms of the new coordinates:    
\begin{eqnarray}
0 & = & \tfrac{1}{2} \frac{\partial^2}{\partial x^2} \left[ x(1 - x)f \right] \nonumber \\
&& + \,  \tfrac{1}{2} \frac{\partial^2}{\partial y^2} \left[ y(1 - y)f \right] \nonumber \\
& & - \,  \tfrac{1}{8} \frac{\partial^2}{\partial x^2} \left[ yf \right] \nonumber \\
& & + \,  \tfrac{1}{2} \frac{\partial}{\partial x}\frac{\partial}{\partial y} \left[(1 - 2x)yf \right] \nonumber \\
& & + \,  \theta \frac{\partial}{\partial x} \left\{ \left[ (q_{12} + \tfrac{1}{2} q_{13}) x - (q_{21} + \tfrac{1}{2} q_{23})(1 - x) \right] f \right\} \nonumber \\
& & + \,  \tfrac{1}{2}\theta \frac{\partial}{\partial x} \left\{ (-\,q_{12} + q_{21} - \tfrac{1}{2} q_{13} + \tfrac{1}{2} q_{23} - q_{31} + q_{32}) yf \right\} \nonumber \\ 
& & - \,  \theta {\frac{\partial}{\partial y}} \left\{ \left[ q_{13} x + q_{23}(1 - x) \right] f \right\} \nonumber \\
& & + \,  \theta {\frac{\partial}{\partial y}} \left\{ (\tfrac{1}{2} q_{13} + \tfrac{1}{2} q_{23} + q_{31} + q_{32} ) yf \right\} \nonumber \\
& = & \mbox{Term 1} + \mbox{Term 2} + \ldots + \mbox{Term 8}.  	\label{XYForwardKolmogorov}
\end{eqnarray}

Guided by the form of the Dirichlet solution Eq.~(\ref{DirSolution}) relevant to the special `parent-independent' case, consider the Ansatz 
\begin{equation} 
f(x, y) = \theta^2 s(x) y^{\theta s(x) - 1} g(x, y).   \label{fAnsatz}
\end{equation} 
Here $s(x)$ and $g(x, y)$ are finite, analytic functions in the region defined by Eq.~(\ref{XYRegion}). The reason for the normalising factor 
$\theta^2$ will become apparent later.  We further expand $g(x, y)$ as the infinite sum
\begin{equation}
g(x, y) = \sum_{k = 0}^\infty g_k(x) y^k,  
\end{equation}
where each $g_k(x)$ is analytic on the $x$-interval in Eq.~(\ref{XYRegion}). 

\begin{table}[t]
\caption{Asymptotic behaviour of each term in Eq.~(\ref{XYForwardKolmogorov}).}
\begin{center}
\begin{tabular}{lll}
\hline
\rule[-2mm]{0mm}{6mm}  & $\lim_{y\rightarrow 0}$(Term $i$) & $\lim_{\theta \rightarrow 0}\int_0^\Lambda (\mbox{Term }i) dy$ \\
\hline
\rule{0mm}{5mm} Term 1:	& $O(y^{\theta s(x) - 1} (\log y)^2)$ & $\frac{1}{2} \theta {d_x}^2[x(1 - x)g_0(x)] + O(\theta^2)$ \\
\rule{0mm}{5mm} Term 2:	& $O(y^{\theta s(x) - 2})$ & divergent \\
\rule{0mm}{5mm} Term 3:	& $O(y^{\theta s(x)} (\log y)^2)$ & $O(\theta^2)$ \\
\rule{0mm}{5mm} Term 4:	& $O(y^{\theta s(x) - 1} \log y)$ & $O(\theta^2)$ \\
\rule{0mm}{5mm} Term 5:	& $O(y^{\theta s(x) - 1} \log y)$ & $O(\theta^2)$ \\
\rule{0mm}{5mm} Term 6:	& $O(y^{\theta s(x)} \log y)$ & $O(\theta^3)$ \\
\rule{0mm}{5mm} Term 7:	& $O(y^{\theta s(x) - 2})$ & divergent \\
\rule{0mm}{5mm} Term 8:	& $O(y^{\theta s(x) - 1})$ & $O(\theta^3)$ \\
\rule{0mm}{5mm} Term 2 $+$ Term 7:&		       & $O(\theta^2)$ \\
\hline
\end{tabular}
\end{center}
\label{tab:termAsymptotes}
\end{table}%

For fixed $x$ and $\theta$ the asymptotic behaviour of each term in Eq.~(\ref{XYForwardKolmogorov}) as $y \rightarrow 0$ is as listed in the first column of 
Table~\ref{tab:termAsymptotes}.  
The dominant terms are Term~2 and Term~7.  Keeping only the dominant $O(y^{\theta s(x) - 2})$ parts of these two terms, Eq.~(\ref{XYForwardKolmogorov}) 
entails that 
\begin{eqnarray}
0 & = & \tfrac{1}{2}s(x) \frac{\partial^2}{\partial y^2} \left[y^{\theta s(x)}\right] - \theta \frac{\partial}{\partial y} \left[ (q_{13} x + q_{23}(1 - x) ) y^{\theta s(x) - 1} \right ] 
																		\nonumber \\
& = & \frac{\partial}{\partial y} \left\{ \tfrac{1}{2}\theta \left[ s(x) - 2(q_{13} x + q_{23}(1 - x) ) \right] y^{\theta s(x) - 1}  \right\} .
\end{eqnarray}
The term inside the curly brackets is the flux of probability across the edge $y = 0$ at a point $x$. This flux must be zero, so 
\begin{equation}
s(x) = 2(q_{13} x + q_{23}(1 - x) ).  \label{sOfX}
\end{equation}

Now define $f_{12}(x)\, dx$ to be the probability contained in the region $[x, x + dx]\times[0, \Lambda]$.   Then 
\begin{eqnarray}
f_{12}(x) & = & \int_0^\Lambda f(x, y)\, dy \nonumber \\
& = & \theta^2 s(x) \sum_{k = 0}^\infty g_k(x) \int_0^\Lambda y^{\theta s(x) - 1 + k} dy \nonumber \\
& = & \theta g_0(x) \Lambda^{\theta s(x)} + \theta^2 \sum_{k=1}^\infty \frac{\Lambda^{\theta s(x) + k}}{\theta s(x) + k} \nonumber \\
& = & \theta g_0(x) + O(\theta^2), \qquad \mbox{as } \theta \rightarrow 0.  \label{IntOfF}
\end{eqnarray}
Similarly we have that 
\begin{equation} 
\int_0^\Lambda yf(x, y)\, dy = O(\theta^2), \quad \int_0^\Lambda \partial_y [y f(x, y)]\, dy = O(\theta^2), \qquad \mbox{as } \theta \rightarrow 0. \label{otherIntTerms}
\end{equation} 
Thus the asymptotic behaviour of the integral of each term in Eq.~(\ref{XYForwardKolmogorov}) is as listed in the second column of 
Table~\ref{tab:termAsymptotes}.  Note that, while the integrals of Term~2 and Term~7 are separately divergent, Eq.~(\ref{sOfX}) ensures that 
there is no $g_0$ contribution to the sum of these terms, which, by a similar calculation to that leading to Eq.~(\ref{otherIntTerms}), integrates to a 
contribution of order $\theta^2$.  Integrating Eq.~(\ref{XYForwardKolmogorov}) term-by-term, dividing by $\theta$, and taking the limit $\theta \rightarrow 0$ 
then gives 
\begin{equation} 
\frac{d^2}{dx^2} \left[ x(1 - x) g_0(x) \right] = 0, 
\end{equation}
whose general solution is 
\begin{equation}
g_0(x) = \frac{c}{x(1 - x)} - \phi\left(\frac{1}{x} - \frac{1}{1 - x}\right), 	\label{g0Soln}
\end{equation}
where $c$ and $\phi$ are arbitrary constants.  Finally, Eq.~(\ref{IntOfF}) implies that $f_{12}(x)$ takes the form of Eq.~(\ref{approx2AlleleSoln}) with constants 
\begin{equation}
C = \theta c, \qquad \Phi = \theta \phi.   	\label{CAndPhi}
\end{equation}
In Section~\ref{sec:K3Case}, $C$ is identified by matching the sum of solutions along neighbouring edges with the normalisation of the effective 2-allele problem 
corresponding to partitioning of alleles, as in \ref{sec:Equiv2}, and $\Phi$ is identified as the flux across a line running from $(x, 0)$ to $(x, \Lambda)$.  
Note from Eq.~(\ref{IntOfF}) that $f_{12}(x)$ is independent of $\Lambda$ provided $\theta \left| \log \Lambda \right| << 1$, or equivalently, 
$\Lambda >>  e^{-(1/\theta)}$.  Note also from Eq.~(\ref{altLineDensityKEquals3}) that $C$ and $\Phi$ are proportional to the rate matrix elements, and 
therefore of order $\theta$.  This justifies the normalisation factor $\theta^2$ in our Ansatz Eq.~(\ref{fAnsatz}).  

\begin{figure}[t]
   \centering
   \includegraphics[width=0.7\textwidth]{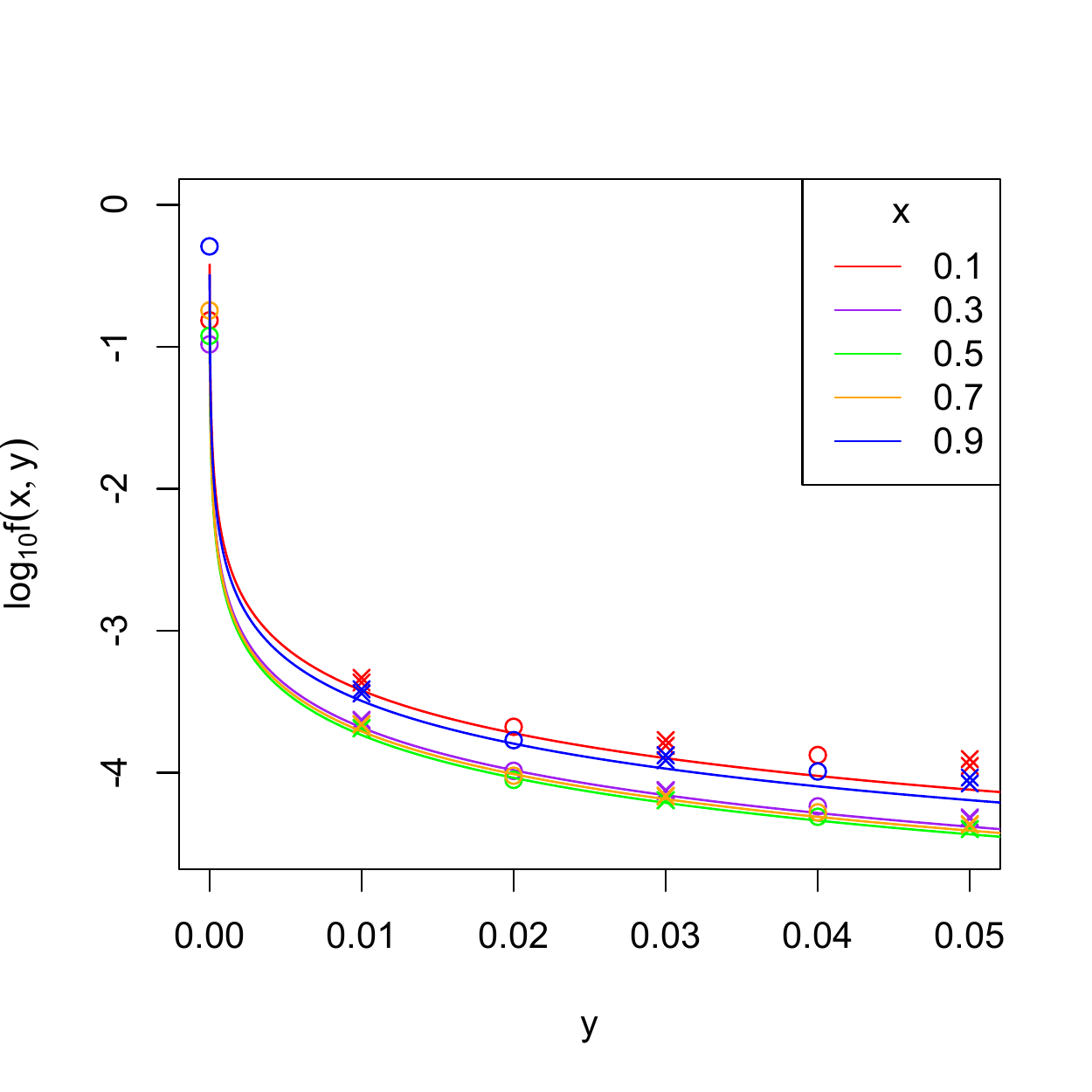} 
   \caption{Comparison between the Ansatz $f(x, y)$ and numerical simulation.  The solid lines are the function $f(x, y)$ defined by Eqs.~(\ref{fAnsatz}), 
   (\ref{sOfX}), (\ref{g0Soln}), (\ref{CAndPhi}) and (\ref{CPhiDefKEquals3}) corresponding to the same rate matrix as Fig.~\ref{fig:3allelesPlot}.  The 
   circles and crosses are a numerical determination of the stationary distribution of the corresponding discrete Wright-Fisher model, Eq.~(\ref{eq:FullPij}), for 
   a population size $N = 100$.  Probabilities of the discrete distribution have been multiplied by $N^2$ for comparison with the continuum probability 
   density.  Circles are simplicial lattice points corresponding to continuum coordinates $(x, y)$, crosses correspond to continuum coordinates 
   $(x \pm (2N)^{-1}, y)$. }  
   \label{fig:TraverseThrough3D}
\end{figure}

Figure~\ref{fig:TraverseThrough3D} is a comparison between the asymptotic analytic solution $f(x, y)$ and a numerical simulation of the stationary 
distribution of the discrete Wright-Fisher model for a population size $N = 100$.  The population size is chosen to be large enough to demonstrate 
the comparison over the first few lattice spacings away from the $A_1$-$A_2$ edge of the simplicial lattice.


\section{Equivalent 2-allele model}
\label{sec:Equiv2}

Suppose we divide the set of allele indices  ${\cal I} = \{1, \ldots , K\}$ into two distinct complementary subsets ${\cal S}_1$ and ${\cal S}_2$, such that 
\begin{equation}
{\cal S}_1 \cup {\cal S}_2 = {\cal I}, \qquad {\cal S}_1 \cap {\cal S}_2 = \emptyset.
\end{equation}
Given an instantaneous $K \times K$ rate matrix $Q$, we wish to construct an effective $2 \times 2$ rate matrix $\tilde Q$ which is equivalent in the sense that 
it defines a 2-state Markov chain whose stationary state is the corresponding marginal distribution of the stationary state of $Q$.  More to the point, 
the transition frequencies between the two states of the effective Markov chain should match the transition frequencies between the two subsets of 
${\cal I}$ in the original Markov chain.  

Suppose $Q$ has elements $Q_{ab}$ for $a, b \in {\cal I}$; $\tilde Q$ has elements $\tilde Q_{AB}$, for $A, B \in\{1, 2\}$; and $L(t)$ is a random variable taking 
values in ${\cal I}$ representing the index of the allele occupied by the original Markov chain at time $t$.  Then the probability of a transition from state $A$ at time $t$ to state $B$ at time $t + \delta t$ is  
\begin{eqnarray}
\delta_{AB} + \tilde Q_{AB} \delta t & = & \Prob\left( L(t + \delta t) \in S_B \right. \left| L(t) \in S_A \right) \nonumber \\ \nonumber \\
	& = & \frac{\sum_{b \in S_B, a \in S_A} \Prob\{L(t + \delta t) = b, L(t) =a \}}{ \sum_{a \in S_A} \Prob\{L(t) =a \}} \nonumber \\ \nonumber \\
	& = & \frac{\sum_{b \in S_B, a \in S_A} (\delta_{ab} + Q_{ab}\delta t) \Prob\{L(t) =a \}}{ \sum_{a \in S_A} \Prob\{L(t) =a \}}.  \label{qTildeDerivation}
\end{eqnarray}
It is clear that there can be no dynamically equivalent 2-state Markov chain in general as the right hand side depends on $t$.  However, as we are only 
interested in equivalence of the stationary behaviour, we can set $t = \infty$ and replace $\Prob\{L(t) =a \}$ with the stationary distribution $\pi_a$, which satisfies 
$\sum_{a = 1}^K \pi_a Q_{ab} = 0$.  This gives 
\begin{eqnarray}
\delta_{AB} + \tilde Q_{AB} \delta t & = & \frac{\sum_{b \in S_B, a \in S_A} (\delta_{ab} + Q_{ab}\delta t)\pi_a}{ \sum_{a \in S_A} \pi_a} \nonumber \\ \nonumber \\
	& = & \frac{\delta_{AB}\sum_{b \in S_B} \pi_b + \sum_{b \in S_B, a \in S_A} \pi_a Q_{ab} \delta t}{ \sum_{a \in S_A} \pi_a} \nonumber \\ \nonumber \\
	& = & \delta_{AB} + \frac{\sum_{b \in S_B, a \in S_A} \pi_a Q_{ab} }{ \sum_{a \in S_A} \pi_a} \delta t.
\end{eqnarray}
Thus the rate matrix of the the equivalent 2-state Markov chain has elements 
\begin{equation}
\tilde Q_{AB} = \frac{\sum_{b \in S_B, a \in S_A} \pi_a Q_{ab}}{ \sum_{a \in S_A} \pi_a}.   \label{equiv2AlleleQ}
\end{equation}
It is easy to see that 
\begin{equation}
\tilde{\pi} _A = \sum_{a \in S_A} \pi_a   \label{equiv2AlleleStat}
\end{equation}
is the corresponding stationary distribution.  

In this paper we have used a conjecture that the marginal distributions of the stationary solution to the forward Kolmogorov equation for a $K$-allele neutral 
Wright-Fisher model corresponding to partitionings of alleles are equal to solutions to the equivalent 2-allele model with a rate matrix $\tilde Q$ whose elements 
are given by Eq.~(\ref{equiv2AlleleQ}).  We have been unable to provide a mathematical proof of this in general, however the result follows easily in the 
restrictive case of the 
solvable $K$-allele model with $Q_{ab} = Q_b$, for which the full stationary distribution is the Dirichlet distribution Eq.~(\ref{DirSolution}).  For this case 
the elements of the rate matrix are of the form $Q_{ab} = \alpha \pi_b$ where $\alpha$ is any real positive constant and $\pi_1, \ldots, \pi_K$ is the stationary 
state of $Q$ normalised so that its elements sum to 1.  Then it follows from Eqs.(\ref{equiv2AlleleQ}) and (\ref{equiv2AlleleStat}) that the equivalent 2-allele 
rate matrix is 
\begin{equation}
\tilde Q_{AB} = \alpha \tilde\pi_B, 
\end{equation}
whose stationary distribution is the beta distribution 
\begin{equation}
\frac{1}{B(2\alpha \tilde\pi_1, 2\alpha \tilde\pi_2)} x^{2\alpha \tilde\pi_1 -1} (1 - x)^{2\alpha \tilde\pi_2 -1}.  
\end{equation}
But using the aggregation property of the Dirichlet distribution~\cite{frigyik2010introduction}, this is precisely the required marginal distribution of the full $K$-allele model, 
Eq.~(\ref{DirSolution}).  


\pagebreak
\section*{References}
\bibliographystyle{elsarticle-num} 
\bibliography{PopulationGenetics}

\end{document}